\documentclass[10pt,openany,a4paper]{article}
\usepackage{geometry,color,framed}
\usepackage{amsmath,amssymb,mathtools,slashed,xcolor,mdframed}
\usepackage[vcentermath,enableskew]{youngtab}
\usepackage{cancel, pifont,multirow}
\usepackage{tikz}
\usetikzlibrary{positioning}
\usetikzlibrary{decorations.text}
\usepackage{amstext}
\usepackage{graphicx}
\usepackage[section]{placeins}
\usepackage{verbatim}
\usepackage{bm}
\usepackage{caption, subcaption, float}
\usepackage[colorlinks=true, 
            linkcolor=blue, citecolor=blue,
            menucolor = red,
            urlcolor=blue,linktoc=all]{hyperref}
\makeatletter

\newcommand{\beq}{\begin{eqnarray}}
\newcommand{\eeq}{\end{eqnarray}}
\newcommand{\non}{\nonumber\\}

\newcommand{\mb}{\mathbb}
\newcommand{\mf}{\mathbf}

\newcommand{\ben}{\begin{enumerate}}
\newcommand{\bei}{\begin{itemize}}
\newcommand{\eni}{\end{itemize}}
\newcommand{\enn}{\end{enumerate}}
\newcommand{\ra}{\rightarrow}
\newcommand{\rr}{\right}
\newcommand{\lf}{\left}

\newcommand{\xdownarrow}[1]{%
  {\left\downarrow\vbox to #1{}\right.\kern-\nulldelimiterspace}
}

\def\l{{\lambda}}

\def\s{{\sigma}}

\usepackage{tikz}
\usetikzlibrary{matrix,shapes,arrows,positioning,chains}
\tikzset{
	 connector/.style={
        -latex,
        font=\scriptsize
    },
    rectangle connector/.style={
        connector,
        to path={(\tikztostart) -- ++(#1,0pt) \tikztonodes |- (\tikztotarget) },
        pos=0.5
    },
    rectangle connector/.default=-2cm,
    straight connector/.style={
        connector,
        to path=--(\tikztotarget) \tikztonodes
    }
}
\makeatother
\title{New non-Euclidean neural quantum states from hyperbolic Lorentz recurrent architectures}
\author{H. L. Dao\footnote{espoirdujour1162@gmail.com}}
\date{\today}

\begin{document} 
\maketitle
\begin{abstract}
In this work, we construct new non-Euclidean neural quantum states (NQS) based on hyperbolic Lorentz recurrent architectures (RNN/GRU). These constructions, together with the Poincar\'e RNN NQS also newly constructed here, extend the class of previously introduced non-Eucllidean NQS which consists only of Poincar\'e hyperbolic GRU. Using the Heisenberg $J_1J_2$ and $J_1J_2J_3$ models consisting of 100 spins in the Variational Monte Carlo (VMC) setting, we show that the four hyperbolic RNN/GRU NQS variants are always able to furnish better representations of the ground state wavefunctions of the quantum systems than their respective Euclidean counterparts with the same architecture. In our experiments, among the four hyperbolic NQS, Lorentz RNN stands out in particular because despite having almost three times fewer parameters, it is capable of surpassing the more complex Poincar\'e GRU and Lorentz GRU to emerge as the best overall hyperbolic NQS ansatz on many instances involving different $J_2$ and ($J_2, J_3$) couplings. Given the findings from this work showing that the four newly constructed hyperbolic RNN/GRU NQS ans\"atze are able to outperform the well-established Euclidean RNN/GRU NQS in Heisenberg spin models, we establish the utility and efficiency of the hyperbolic Lorentz RNN/GRU NQS as well as the  Poincar\'e RNN/GRU NQS for future variational studies of quantum many-body systems, especially those exhibiting a hierarchical structure in the form of the different degrees of nearest-neighbor interactions.
\end{abstract}

\tableofcontents
\clearpage
\section{Introduction}
The field of neural quantum states (NQS) that utilizes neural networks to approximate ground state wavefunctions of quantum many-body systems, most notably spin systems such as the Transverse Field Ising Model (TFIM) and Heisenberg $J_1J_2$ models, have been expanding rapidly since the introduction of the Restricted Boltzmann Machine (RBM) as the first type of NQS in the seminal works \cite{1606-rbm}, \cite{1610-rbm}. Since then, many different types of NQS based on different neural network architectures such as simple artificial neural networks (ANN) (\cite{1704-ann}, \cite{1709-ann}), convolutional neural networks (CNNs) (\cite{cnn-1807}, \cite{cnn-1903}, \cite{gcnn-2211}), recurrent neural networks (RNNs) (\cite{rnn_20}, \cite{rnn_22}, \cite{rnn-24}, \cite{rnn-25}), and transformers (\cite{2211-transformer}, \cite{transformer-2311}, \cite{2306-transformer}, \cite{2406-transformer}, \cite{transformer-25}) - to name a few - have been developed in rapid succession. While these works explored the various advantages that the many different neural network architectures had to offer and showed that they could deliver impressive results, the one single constant among them all is the fact that they are all conventional Euclidean constructions. In other words, the mathematical operations underlying their constructions are defined on Euclidean space. It was not until \cite{hld-hypnqs-25} that the first type of non-Euclidean NQS ansatz was introduced in the form of the Poincar\'e hyperbolic GRU (Gated Recurrent Unit \cite{cho-gru} - a specialized form of RNNs), whose underlying construction made use of the geometry of the Poincar\'e disk model of hyperbolic space, based on the prior work \cite{ganea-1805} in the field of Natural Language Processing (NLP) that first introduced the hyperbolic Poincar\'e RNN/GRU networks for the NLP tasks  of sentence embeddings, textual entailment and noisy-prefix recognition. In NLP, later works on hyperbolic neural networks \cite{hyprnn-lorentz-2105}-\cite{1911-cnn} have consistently shown that they are capable of outperforming their Euclidean counterparts, especially when the involved datasets have hierarchical, tree-like structures. This has to do with the fact that hyperbolic space can embed tree-like structures with very low distortions \cite{sarkar-11}, thanks to the property that hyperbolic geometry allows for an exponential expansion of space (area or volume) as a function of the radial distance, unlike Euclidean space, which grows only polynomially with distance \cite{krioukov-09}, \cite{krioukov-10}, \cite{sala-2018}.
\\\\
As a proof of concept, in \cite{hld-hypnqs-25}, it was shown that the then-newly introduced hyperbolic NQS ansatz based on Poincar\'e GRU (referred to simply as `hyperbolic GRU' in \cite{hld-hypnqs-25}) could definitively outperform the ordinary NQS ans\"atze based on Euclidean GRU and RNN in the three settings of the two-dimensional TFIM of up to 81 spins, the one-dimensional Heisenberg $J_1J_2$ model with 50 spins, and the one-dimensional Heisenberg $J_1J_2J_3$ model with 30 spins. In particular, the outperformance delivered by Poincar\'e GRU was definitive when the systems under consideration exhibited a clear hierarchical structures, such as the three cases mentioned above which involved different competing degrees of nearest neighbor interactions. It was hypothesized in \cite{hld-hypnqs-25} that an NQS ansatz based on hyperbolic GRU would likely outperform one based on Euclidean GRU in all such quantum many-body systems, including those already considered but with much larger system sizes. Furthermore, the work \cite{hld-hypnqs-25} also hypothesized the possibility of potentially similar outperformances delivered by other types of hyperbolic NQS ans\"atze over their Euclidean versions. While these yet-to-be-constructed hyperbolic NQS ans\"atze could be based on any kind of different architectures, such as CNNs, RBMs or transformers, the most immediate and natural extension of \cite{hld-hypnqs-25} would arguably be the construction of the hyperbolic version of the simpler RNN architecture, which was not done in \cite{hld-hypnqs-25}, as well as the construction of both the RNN and GRU architectures using the Lorentz model of hyperbolic space, which is believed to have the potential to outperform the Poincar\'e hyperbolic model with the same architecture.
\\\\
In this work, our main motivations and results are summarized below.
\bei
\item As a first step, we constructed and introduced three new types of hyperbolic RNN NQS ans\"atze: Poincar\'e RNN, Lorentz RNN and Lorentz GRU, alongside the Poincar\'e GRU\footnote{The four variants will be collectively referred to as hyperbolic RNNs.} originally introduced in our previous work \cite{hld-hypnqs-25}. Incidentally, to the best of our knowledge, the Lorentz RNN and Lorentz GRU in this work are the first standalone constructions of their kinds in the literature, i.e. the first constructions where RNN/GRU architectures are used in conjunction with the Lorentz geometry. In the NLP literature where hyperbolic neural networks originated, only the Poincar\'e GRU/RNN were explicitly formulated in the works of \cite{ganea-1805}, and while many different neural network architectures  based on Lorentz geometry incorporating some forms of recurrent gates were explored in later NLP works (such as \cite{lrnn-s1}), none explicitly constructed the Lorentz RNN or Lorentz GRU as standalone, independent architectures. More specifically, while Lorentz embeddings \cite{hyprnn-lorentz-1806}, Lorentz feed-forward layers, and Lorentz attention \cite{hyprnn-lorentz-2105} were widely used as a direct result of the transition from Poincar\'e geometry to Lorentz geometry in the NLP context, a standalone Lorentz RNN/GRU gating mechanism was never formally formulated. The recurrent gating equations operating natively on the Lorentz hyperboloid are only explicitly derived and benchmarked in this work in the context of many-body quantum physics.
\item
Secondly, using the settings of the one-dimensional Heisenberg $J_1J_2$ and $J_1J_2J_3$ models with 100 spins, which is two times larger than the $J_1J_2$ model and three times larger than the $J_1J_2J_3$ model considered in \cite{hld-hypnqs-25}, we carried out the variational Monte-Carlo (VMC) experiments using six types of NQS ans\"atze: Euclidean RNN/GRU, Poincar\'e RNN/GRU, Lorentz RNN/GRU to benchmark the performances of all four hyperbolic RNNs against their Euclidean counterparts. As first raised in \cite{hld-hypnqs-25}, the primary question to settle in this work is still whether, in the specific larger-size Heisenberg quantum many-body systems under study,  these collective hyperbolic RNN NQS can outperform the Euclidean NQS with the same architecture. However, with the inclusion of the new types of hyperbolic RNNs, additional questions also arise regarding the performances of different NQS ans\"atze constructed from different models or representations of hyperbolic space relative to each other. More specifically, the secondary question to explore in this work is, which hyperbolic sub-model, Lorentz or Poincar\'e, will definitively proves to be better.
\\\\
While the primary  question regarding the outperformance of the collective hyperbolic RNN NQS over the Euclidean NQS can be affirmatively confirmed, the secondary question is less definitively so, since a different hyperbolic model emerged as the top performing NQS in a  different coupling regime of the Heisenberg mdel under study. This perhaps has less to do with the efficiency or effectiveness of the NQS models themselves than with the physics of the problem, more specifically, it makes sense that with different underlying physics (represented by the small or large coupling constants $J_2$ and $(J_2,J_3)$ in the system), no single hyperbolic NQS ansatz can always emerge as the best approximation to the true ground state every single time. Instead, what we would see from the results of this work is that on the majority of the occasions, out of the four new hyperbolic RNN variants, Lorentz RNN emerged as the top performing variant, surpassing both Poincar\'e GRU and Lorentz GRU, while on another occasion, Poincar\'e GRU was the top performer, and on the remaining occasion, Lorentz GRU took the top spot. The emergence of the Lorentz RNN as the overall best hyperbolic NQS variant, despite having three times fewer parameters than Lorentz/Poincar\'e GRU (and Euclidean GRU), shows that it is a parameter-efficient construction that is valuable to future VMC studies of all kinds of quantum many-body systems beyond spin models.
\eni
This paper is orgarnized as follows: In Section \ref{sec-phyprnn}, we detail the mathematical constructions of the Poincar\'e RNN and GRU (Section \ref{phrnn}), as well as the constructions of the Lorentz RNN and GRU (Section \ref{sec-lhyprnn}). In Section \ref{sec-vmc-summary}, we use these newly constructed hyperbolic RNNs to build new non-Euclidean, hyperbolic NQS ans\"atze. In Section \ref{sec-1dj1j2} and \ref{sec-1dj1j2j3}, we report and discuss the results of the VMC experiments using the six types of (hyperbolic and Euclidean) NQS ans\"atze in the Heisenberg $J_1J_2$ and Heisenberg $J_1J_2J_3$ model, respectively. In Section \ref{sum-concl}, we summarize the findings of this work and describe some possible future directions. In the Appendix, Section \ref{spatial-constr} discusses the effects of spatial constraint hyperparameters on the performances of the Lorentz and Poincar\'e hyperbolic neural networks, Sections \ref{full-res-j1j2} and \ref{full-res-j1j2j3} include the full results of all NQS ans\"atze from different VMC runs using different random seeds.
\\\\
The Python codes used in this work to construct the hyperbolic Poincar\'e RNN/GRU NQS are the \texttt{Pytorch} translation of the \texttt{TensorFlow} codes\footnote{which can be found at the Github directory: \href{https://github.com/lorrespz/nqs_hyperbolic_rnn}{https://github.com/lorrespz/nqs\_hyperbolic\_rnn} } written for our earlier work \cite{hld-hypnqs-25}, while the codes used to construct the Lorentz RNN/GRU NQS are built from the Poincar\'e version, but the actual construction of the Lorentz RNN/GRU hyperbolic networks,  which formed the main component of the Lorentz-based NQS ans\"atze, are based on the \texttt{hypercore} library \cite{hypercore} with some modifications. In particular, we used the files \texttt{lmath.py} and \texttt{lorentzian.py} from the subdirectory \texttt{manifolds} of the main directory \texttt{hypercore\_main} to construct the Lorentz RNN and Lorentz GRU\footnote{The library \texttt{hypercore} offers a comprehensive toolbox for Lorentz-based hyperbolic models, and comes with many full-fledged hyperbolic networks that are preconstructed by the authors such as Lorentz MLP, Lorentz CNN, Lorentz Graph Neural Network, and Lorentz transformers. However, the RNN-based family of models containing RNN, GRU and LSTM (Long Short Term Memory) are not among these preconstructed models. The Lorentz RNN and GRU are constructed for this first time as independend, standalone architectures in this work. }.
All Python codes and trained neural networks for this work can be found at this Github repository: \href{https://github.com/lorrespz/hypnqs_lorentz_poincare}{https://github.com/lorrespz/hypnqs\_lorentz\_poincare}.

\section{Two variants of hyperbolic RNNs}\label{sec-phyprnn}
The constructions of the Poincar\'e and Lorentz variants of hyperbolic RNNs are based on the following definitions of Euclidean RNN and GRU. 
Starting with an input vector $\vec x_i \in \mb{R}^{d_x}$ at step $i$ of size $d_x$, the conventional version of RNN is a function that relates the hidden state vector $\vec h_i\in \mb{R}^{d_h}$ of size $d_h$ to the input $\vec x_i$ and the same hidden state vector $\vec h_{i-1}$ at the previous time step $(i-1)$
\beq
\text{RNN} : && \vec h_i = f(W_h \vec h_{i-1} + U_h\vec x_{i} + \vec b_h)\,, \label{eq-rnn}
\eeq
where $f$ is a nonlinear activation function (often `tanh'), $W_h$ is a $d_h\times d_h$ weight matrix, $U_h$ is a $d_h\times d_x$ weight matrix, and $b_h\in \mb{R}^{d_h}$ is a vector of size $d_h$ known as the bias.
\\\\
GRU \cite{cho-gru} is a more sophisticated version of RNN with additional structures comprising the reset gate $\vec r_i \in \mb R^{d_h}$ and the update gate $\vec z_i\in \mb R^{d_h}$. The defining equations of the conventional or Euclidean GRU are
\beq
\text{GRU}: &&   \vec r_i = \s\lf(W_r   \vec h_{i-1} + U_r \vec x_i +   \vec b_r\rr) \non 
            &&   \vec z_i = \s\lf(W_z   \vec h_{i-1} + U_z \vec x_i +   \vec b_z\rr) \non
            &&   \vec{\tilde h}_i = f\lf[W_h(  \vec r_i \odot  \vec h_{i-1}) + U_h\vec x_i +   \vec b_h\rr] \non
 &&   \vec h_i = (1-\vec z_i)\odot   \vec h_{i-1} +   \vec z_i\odot   \vec{\tilde h}_i \label{gru-update}
\eeq
In Eq.(\ref{gru-update}), $\vec h_i$ is the final hidden state at step $i$ while $\vec{\tilde h}_i$ is the new state computed in the same time step, $f$ is a nonlinear activation function (often taken to be `tanh'), $\s$ is the sigmoid activation, $\vec x_{i}$ is the input vector of length $d_x$, $W_h, W_z, W_r$ are the $d_h\times d_h$ weight matrices, $U_h, U_z, U_r$ are the $d_h\times d_x$ weight matrices, $\vec b_h,   \vec b_z,   \vec b_r \in \mathbb{R}^{d_h}$ are the bias vectors, and $\odot$ is the pointwise multiplication operation.
Note that for GRU, the last equation appearing in Eq.(\ref{gru-update}) that defines the final hidden state $\vec h_i$ can also be written as follows
\beq
\vec h_i = \vec h_{i-1} + \vec z_i\odot(-\vec h_{i-1}+ \vec{\tilde h}_i)\,, \label{eq:gru-update-v2}
\eeq
 where the update step involves the previous state $\vec h_{i-1}$ and the new state $\vec{\tilde h}_i$, which is a function of the reset gate $\vec r_i$, the previous state $\vec h_{i-1}$ and the input $\vec x_i$. When $\vec z_i \approx 0$, the final state $\vec h_i$ is almost entirely $\vec{\tilde h}_i$ - the new state. When $\vec z_i \approx 1$, the final state $\vec h_i$ is almost entirely its previous state $\vec h_{i-1}$. When $\vec r_i \approx 1$ and $\vec z_i \approx 1$, the GRU essentially reduces to the RNN \cite{hld-cicy-rnn}.
\subsection{Construction of Poincare hyperbolic RNN/GRU} \label{phrnn}
The Poincar\'e ball model $(\mathbb{D}^N, g^{\mathbb D})$ of hyperbolic space is defined by the manifold $\mathbb{D}^N$
\beq
\mathbb{D}^N_c = \left\{ x\in \mathbb{R}^N: c||x||<1\right\} \label{eq-gyro}
\eeq
where the parameter $c$ is the Poincar\'e ball's radius. When $c=0$, $\mathbb{D}^N_c = \mathbb{R}^N$, while when $c>0$, $\mathbb{D}^N_c$ is the open ball of radius $1/\sqrt{c}$. In all the computations used in this work, we set $c=1$. 
The space $\mathbb{D}^N_c$ in Eq.(\ref{eq-gyro}) is equipped with the metric
\beq
g^{\mb D}_x = \lambda^2_x g^E, \hspace{5mm} \lambda_x = \frac{2}{1-||x||^2} \label{eq:poincare-metric}
\eeq
where $g^E = \mathbf{1}_N$ is the identity matrix representing the Euclidean metric. 
The parallel transport $P^c_{\mf{0}\ra x}(v)$ of a vector $v\in T_\mf{0}\mb{D}^N_c$ to another tangent space $T_x\mb{D}^N_c$ are defined as
\beq
P^c_{\mf{0}\ra x}(v) = \log_x^c\lf[x\oplus_c \exp_\mf{0}^c(v)\rr] \label{eq-pt}
\eeq
where the operations $\log$ and $\exp$ are the exponential and logarithmic maps. For any point $x\in \mb{D}^N_c$, any vector $v\neq \mf{0}$ and any point $y\neq x$, the exponential and logarithmic maps $\exp^c_x: T_x\mb{D}^M_c \ra \mb{D}^M_c $ and $\log^c_x: \mb{D}^N_c \ra T_x\mb{D}^N_c$ between the hyperbolic space and its Euclidean tangent space are defined as
\beq
\exp_x^c(v) &=& x \oplus_c\lf[ \tanh\lf(\sqrt{c}\frac{\lambda_x^c||v||}{2}\rr)\frac{v}{\sqrt{c}||v||}\rr] \label{eq-exp}
\\
\log_x^c(y) &=& \frac{2}{\sqrt{c}\l_x^c} \tanh^{-1}\lf(\sqrt{c}||-x\oplus_c y||\rr) \frac{-x\oplus_cy}{||-x\oplus_c y||} \label{eq-log}
\eeq
where $\l_x^c = 2/(1-c||x||^2)$. When $x = \mf{0}$, the above maps take  more compact forms
\beq
\exp_\mf{0}^c(v) &=& \tanh\lf(\sqrt{c}||v||\rr) \frac{v}{\sqrt{c}||v||}\,\,\qquad 
\lf(T_{\mf{0}_M}\mb{D}^M_c \ra \mb{D}^M_c \rr) \label{eq-exp0}
\\
\log_\mf{0}^c(y) &=& \tanh^{-1}\lf(\sqrt{c}||y||\rr) \frac{y}{\sqrt{c}||y||}\qquad \lf(\mb{D}^N_c \ra T_{\mf{0}_N}\mb{D}^N_c\rr)\,. \label{eq-log0}
\eeq
Poincar\'e hyperbolic RNNs were discussed in detail in \cite{hld-hypnqs-25} but for the sake of completeness, we will summarize their constructions here.
The Poincar\'e hyperbolic RNN and hyperbolic GRU are defined by the following formulae
\beq
\text{Hyperbolic RNN} : && \vec h_i = (W_h \otimes_c \vec h_{i-1}) \oplus_c (U_h\otimes_c \vec x_{i}) \oplus_c \frac{}{}\vec b_h\label{eq-hrnn}
\\  \non
 \text{Hyperbolic GRU} : && 
  \vec r_i = \s\,\lf\{\log_\mf{0}^c\lf[(W_r\otimes_c \vec h_{i-1}) \oplus_c (U_r\otimes_c \vec x_i) \oplus_c \frac{}{}   \vec b_r\rr]\rr\}\,,
 \non
  && \vec z_i =\s\,\lf\{\log_\mf{0}^c\lf[(W_z\otimes_c \vec h_{i-1}) \oplus_c (U_z\otimes_c \vec x_i) \oplus_c \frac{}{}  \vec b_z\rr]\rr\}\,, \label{eq-hgru}
  \\
  && \vec{\tilde h}_i = W_h\,\otimes_c(\vec r_i \odot_c \vec h_{i-1}) \oplus_c (U_h\otimes_c \vec x_i)\frac{}{} \oplus_c   \vec b_h\,, 
  \non
  && \vec h_i =\vec h_{i-1} \oplus_c \,\vec z_i\, \odot_c\lf(-\vec h_{i-1} \oplus_c \vec{\tilde h}_i\rr)\,. \nonumber
 \eeq
In Eqs.(\ref{eq-hrnn}), (\ref{eq-hgru}), the various weight matrices $W_{h,r,z}$ and $U_{h,r,z}$ have the same meanings as in the Euclidean RNN/GRU case - Eqs.(\ref{eq-rnn}), (\ref{gru-update}), but the hidden state vector $\vec h_i \in \mb{D}^{d_h}_c$, input vector $\vec x_i \in \mb{D}^{d_i}_c$, and bias vectors $\vec b_{h,r,z} \in\mb{D}^{d_h}_c$ are all vectors on the Poincar\'e disk. The nonlinear activation function $f$ (often tanh) used in the Euclidean RNN/GRU is not used in the hyperbolic case (in this work) because hyperbolic space itself provides some degree of nonlinearity.
 The mathematical operations $\oplus_c, \otimes_c, \odot_c$ appearing in the equations above are the Poincar\'e hyperbolic analogs of the Euclidean addition, matrix multiplication and pointwise multiplication. Although these formulae were derived in \cite{ganea-1805} and described in more detail in \cite{hld-hypnqs-25}, they are  recalled below for the sake of completeness.
\bei
\item For $x,y \in \mb{D}^N_c$, the Mobius addition $\oplus_c$ is
\beq
x\oplus_c y \equiv \frac{(1+2c\langle x, y\rangle + c||y||^2)x + (1-c||x||^2)y}{1 + 2c \langle x, y\rangle + c^2 ||x||^2 ||y||^2}\,. \label{eq-oplus}
\eeq
\item For $x\in \mathbb{D}^N_c\backslash \{\mathbf{0}\}$ and $W: \mathbb R^M \ra \mathbb R^N$, the Mobius matrix multiplication is defined by
\beq
 W \otimes_c x = \frac{1}{\sqrt{c}} \tanh\lf(\frac{||Wx||}{||x||} \tanh^{-1}(\sqrt{c}||x||)\rr)\frac{Wx}{||Wx||}
\eeq
\item For $x\in \mathbb{D}^N_c\backslash \{\mathbf{0}\}$ and $r\in \mathbb R$, the Mobius pointwise multiplication $\odot_c$, is defined as
\beq
r\odot_c x = \frac{1}{\sqrt{c}} \tanh\lf(\frac{||rx||}{||x||} \tanh^{-1}(\sqrt{c}||x||)\rr)\frac{rx}{||rx||}
\eeq
\eni
In terms of parallel transport, the Mobius addition for $x\in \mb{D}^N_c$ with $b\in \mb{D}^N_c$ can be written as
\beq
x\oplus_c b = \exp_x^c\lf[P_{\mf{0} \ra x}^c\lf(\log_\mf{0}^c(b)\frac{}{}\rr)\rr]\label{eq-m-pt}\,.
\eeq
\subsection{Construction of Lorentz hyperbolic RNN/GRU}\label{sec-lhyprnn}
The $n$-dimensional Lorentz hyperboloid $\mathbb H^n$ with constant negative curvature $-k$ where $k=1$ is given as
\beq
\mathbb H^n \equiv \left\{\mathbf x\in \mathbb R^{n+1}: \langle \mathbf x,\mathbf x\rangle_{\mathcal L} = -1, \,x_0>0\rr\}
\eeq
where $\langle \mathbf x, \mathbf y \rangle_{\mathcal L}$, the Lorentzian scalar product for $(n+1)$-dimensional vectors $\mathbf{x, y} \in \mathbb R^{n+1}$, is defined as
\beq
\langle \mathbf x, \mathbf y \rangle_{\mathcal L} \equiv -x_0y_0 + \sum_{i=1}^n x_iy_i\,.
\eeq
Unlike the Poincar\'e disk where the origin $\mathbf 0_P = (0,0,\ldots, 0)$, in the Lorentz model of hyperbolic space, the origin is the point $\mathbf 0_{\mathcal L} = (1, 0, 0,\ldots, 0)$.
The distance function between two points $\mathbf x, \mathbf y \in \mathbb H^n$ is 
\beq
d_{\mathbb H}(\mathbf x, \mathbf y) = \text{arcosh}\left(-\langle  \mathbf x, \mathbf y\rangle_{\mathcal L}\rr)\,,
\eeq
while the Lorentzian norm of a vector $\mathbf v$ is 
\beq
||\mathbf v||_{\mathcal L} = \sqrt{\langle  \mathbf v, \mathbf v\rangle_{\mathcal L}}\,.
\eeq
The tangent space at $\mathbf x$ is the $n$-dimensional Euclidean vector space approximating $\mathbb H^n$ around $\mathbf x$:
\beq
\mathcal T_x\mathbb H^n \equiv \lf\{ \mathbf x\in \mathbb R^{n+1}: \langle \mathbf v, \mathbf x\rangle_{\mathcal L} = 0\rr\}
\eeq
Similar to the Poincar\'e model described in the previous section, mappings between the Lorentz hyperboloid and its tangent space are done using the exponential $\exp_{\mathbf x}(\mathbf v)$ and logarithmic $\log_{\mathbf{x}}(\mathbf y)$ maps.
\beq
\mathcal T_{\mathbf x} \mathbb H^n \rightarrow \mathbb H^n: &&\exp_{\mathbf x}(\mathbf v) = \cosh\lf(||\mathbf v||_{\mathcal L}\rr) \mathbf x + \sinh\lf(||\mathbf v||_{\mathcal L}\rr)\frac{\mathbf v}{||\mathbf v||_{\mathcal L}} \label{eq-l-expmap}\\
\mathbb H^n \rightarrow \mathcal T_{\mathbf x} \mathbb H^n: && \log_{\mathbf{x}}(\mathbf y) =d_{\mathbb H}(\mathbf x, \mathbf y) \frac{\mathbf y + \langle \mathbf x, \mathbf y\rangle_{\mathcal L}\,\mathbf x}{\left||\mathbf y + \langle \mathbf x, \mathbf y\rangle_{\mathcal L}\,\mathbf x\right||_{\mathcal L}} \label{eq-l-logmap}
\eeq
The parallel transport operation that maps a point $\mathbf{z} \in \mathcal T_{\mathbf x} \mathbb H^n$ to a point in $\mathcal T_{\mathbf y}\mathbb H^n$ is defined as
\beq
P_{\mathbf{x}\ra \mathbf{y}}(\mathbf{z}) = \mathbf z+\frac{\langle \mathbf y, \mathbf z\rangle_{\mathcal L}}{1-\langle(\mathbf x, \mathbf y)\rangle_{\mathcal L}}(\mathbf x+ \mathbf y)
\eeq
When $\mathbf x= \mathbf 0_{\mathcal L}$, 
\beq
P_{\mathbf 0_{\mathcal L} \ra \mathbf{y}}(\mathbf z) = \mathbf z+\frac{\langle \mathbf y, \mathbf z\rangle_{\mathcal L}}{1-\langle(\mathbf 0_{\mathcal L}, \mathbf y)\rangle_{\mathcal L}}(\mathbf 0_{\mathcal L}+ \mathbf y) \label{ptrans0}
\eeq
Using the Euclidean definitions of RNN and GRU in Eqs.(\ref{eq-rnn}), (\ref{gru-update}), the Lorentz version of hyperbolic RNN and GRU (referred to as Lorentz RNN and Lorentz GRU, respectively) are defined as
\beq
\text{Lorentz RNN}: && \vec h_i = \lf(W_h \otimes_\mathcal{L} \vec h_{i-1} \rr) \oplus_\mathcal{L} \lf(U_h \otimes_\mathcal{L} \vec x_i\rr) \oplus_\mathcal{L} \vec b_h \label{eq-lorentz-rnn} \\
\text{Lorentz GRU}: && \vec r_i = \s\lf\{\log_{\mathbf 0_{\mathcal L}}\lf[\lf(W_r \otimes_\mathcal{L} \vec h_{i-1} \rr) \oplus_\mathcal{L} \lf(U_r \otimes_\mathcal{L} \vec x_i\rr) \oplus_\mathcal{L} \vec b_r\rr] \rr\}\non
&& \vec z_i = \s\lf\{\log_{\mathbf 0_{\mathcal L}}\lf[\lf(W_z \otimes_\mathcal{L} \vec h_{i-1} \rr) \oplus_\mathcal{L} \lf(U_z \otimes_\mathcal{L} \vec x_i\rr) \oplus_\mathcal{L} \vec b_z \rr]\rr\}\label{eq-lorentz-gru}
  \\
  && \vec{\tilde h}_i = W_h\,\otimes_{\mathcal L}(\vec r_i \odot_{\mathcal L} \vec h_{i-1}) \oplus_{\mathcal L} (U_h\otimes_{\mathcal L} \vec x_i)\frac{}{} \oplus_{\mathcal L}  \vec b_h\,,
  \non
  && \vec h_i = \vec h_{i-1} \oplus_{\mathcal L} \vec z_i\odot_{\mathcal L}\lf(- \vec h_{i-1} \oplus_{\mathcal L} \vec{\tilde h}_{i}\rr)\,. \nonumber
\eeq
where the weight matrices $W_{h,r,z}$, $U_{h,r,z}$ and bias vectors have the same meaning as previously defined. 
The various Lorentz mathematical operations appearing Eqs.(\ref{eq-lorentz-rnn}), (\ref{eq-lorentz-gru}) are defined in terms of the exponential/logarithm mappings at $\mathbf x= \mathbf 0_{\mathcal L}$ in (\ref{eq-l-expmap}), (\ref{eq-l-logmap}) and parallel transport operations given in (\ref{ptrans0})  as follows.
\bei
\item For $\mathbf x, \mathbf y \in \mathbb H^n$, the Lorentz addition $ \oplus_{\mathcal L}$ is defined as:
\beq
\text{Lorentz addition}: \qquad \mathbf x \oplus_{\mathcal L} \mathbf y = \exp_{\mathbf x}\lf[P_{\mathbf 0_{\mathcal L} \ra \mathbf{x}}(\log_{\mathbf 0_{\mathcal L}}(\mathbf y))\rr]
\eeq
\item The scalar multiplication $\odot_{\mathcal L}$ between a point $\mathbf x\in \mathbb H^n$ and $r\in \mathbb R$ is 
\beq
\text{Lorentz scalar multiplication}: \qquad r \odot_{\mathcal L} \mathbf x =\exp_{\mathbf 0_{\mathcal L}}\lf[ r \log_{\mathbf 0_{\mathcal L}}(\mathbf x)\rr]
\eeq
\item The matrix multiplication $\otimes_{\mathcal L}$ between a point $\mathbf x\in \mathbb H^n$ and a matrix  $M\in \mathbb R^n\times \mathbb R^n$ is 
\beq
M \otimes_{\mathcal L} \mathbf x =\exp_{\mathbf 0_{\mathcal L}}\lf[ M \log_{\mathbf 0_{\mathcal L}}(\mathbf x)\rr]
\eeq
\eni
In Eqs.(\ref{eq-lorentz-rnn}), (\ref{eq-lorentz-gru}), the RNN/GRU hidden state $\vec h_{i}$ at the end of each computation is hyperbolic and lives on the Lorentz hyperboloid.  The weight matrices $W_{h,r,z}, U_{h,r,z}$  are Euclidean, while the biases $\vec b_{h,r,z}$ can be hyperbolic or Euclidean (in which case, they are projected onto the hyperboloid using the exponential map $\exp_{\mathbf{0}_\mathcal L}$). For the Lorentz GRU,  the last equation in Eq.(\ref{eq-lorentz-gru}) is the update step for the hidden state $\vec h_i$, which was done on the hyperboloid in an identical manner to the Poincar\'e case\footnote{We note that it is also possible to define the update step using the tangent space as follows: \beq
  \vec h_i =\vec h_{i-1} \oplus_{\mathcal L} \exp_{\mathbf 0_{\mathcal L}}\lf\{\vec z_i\, *\lf[\log_{\mathbf 0_{\mathcal L}}(\vec{\tilde h}_i) -\log_{\mathbf 0_{\mathcal L}}(\vec h_{i-1})\rr]\rr\}
\eeq
where the hyperbolic hidden states $\vec{\tilde h}_i$ and $\vec h_{i-1}$ have to be projected to the tangent space first before the subtraction is done, and the resulting state is multiplied with $\vec z_i$ in the tangent space before the exponential map that projects the result back on the hyperboloid, to be added to the hyperbolic state $\vec h_{i-1}$. However, this definition relies heavily on the tangent space and is not as elegant as the one above, defined using the manifold primarily.
}. However, it must be noted that in Lorentz hyperbolic space, the state $-\vec h_{i-1}$ is not simply the state $\vec h_{i-1}$ with all components multiplied by $-1$ (as is the case in Euclidean space as well as in the Poincar\'e disk). Instead, the $-\vec h_{i-1}$ state has the components  $(x_0, -\vec x_i)$ if $\vec h_{i-1} = (x_0, \vec x_i)$, since we need $\vec h_{i-1} \oplus_{\mathcal L} (-\vec h_{i-1}) = \mathbf 0_{\mathcal L}$.
\section{Hyperbolic RNN-based NQS wavefunctions}\label{sec-vmc-summary}
In this section, we summarize the main points regarding the variational Monte-Carlo method, as well as the constructions of the real and complex NQS wavefunction using either Euclidean or hyperbolic RNN/GRU in discrete Hamiltonian spin systems. The detailed description can be found in \cite{rnn_20}, \cite{hld-hypnqs-25}.
\\\\
\textit{Variational Monte Carlo (VMC)}: The variational Monte Carlo (VMC) method, often used to train NQS, involves the process of sampling from a probability distribution represented by the square of the trial wavefunction/the variational ansatz and subsequently using these generated samples to calculate some obervables such as the ground state energy.
In this section, we recall the derivation of the local energy formula in the Variational Monte Carlo (VMC) method \cite{mc-textbook}.
\\\\
Given a quantum Hamiltonian $H$, in VMC, the local energy $E_\text{loc}(x)$ of samples $|x\rangle$ generated by an NQS $\Psi$ is given by 
\beq
E_\text{loc}(x) = \sum_{x'} \langle x|H|x'\rangle \frac{\langle x'|\Psi\rangle}{\langle x|\Psi\rangle}. \label{Eloc}
\eeq
$E_\text{loc}(x)$ is non-zero only for those non-zero Hamiltonian elements $\langle x|H|x'\rangle \neq 0$. Corresponding to each generated sample $|x\rangle$ is a probability
\beq
P_\text{loc}(x)= \frac{|\Psi(x)|^2}{\sum_{|x\rangle}|\Psi(x) |^2} \label{Ploc}
\eeq
In the variational problem of interest, given a Hamiltonian $H$ and a trial wavefunction $\Psi$, the VMC task is to estimate the ground state energy $E = \langle \Psi|H|\Psi\rangle$ which can be written in terms of the  local energy $E_\text{loc}$ and the probability $P_\text{loc}(x) $
\beq
E=\sum_{|x\rangle}P_{\text{loc}}(x) E_\text{loc}(x)\,. \label{full-e}
\eeq
When the trial wavefunction is a neural network quantum state $|\Psi(\vec\theta)\rangle$ with trainable parameters $\vec \theta$, at each training iteration $i$, the variational energy $E(\vec\theta)$ is calculated from the local energy $E_\text{loc}(\vec\theta)$ of the Monte Carlo samples generated from the NQS using Eq.(\ref{full-e}). The process of minimizing $E(\vec\theta)$ using an optimizer (such as SGD - Stochastic Gradient Descent or Adam - Adaptive Moment Estimation) updates the trainable parameters $\vec\theta$ until convergence is reached.
\\\\
For the ensuing discussion below regarding the constructions of the real and complex NQS ans\"atze, the basis state of the Hamiltonian $H$ is denoted by $|\vec\s\rangle = (\s_1, \ldots, \s_N)$ where $N$ is the dimensionality of $H$, and each component $\s_i (1\leq i\leq N)$ assumes discrete values of either 0 or 1.
\begin{itemize}
    \item \textbf{Real NQS wavefunction}: The wavefunction is defined as
    \beq
    |\Psi\rangle = \sum_{\vec\s} \sqrt{P(\vec \s)}|\vec \s\rangle
    \eeq
    where $P(\vec\s)$, the probability of a particular configuration $|\vec\s\rangle$, is the output of the real NQS. 
    \beq
    P(\vec \s) = P(\s_1) P(\s_2|\s_1) \ldots P(\s_N|\s_1, \s_2, \ldots, \s_{N-1})
    \eeq
\begin{figure}[H]
\centering
\begin{tikzpicture}[node distance = 1.3cm, thick]%
        \node[circle, draw] (0) {$P(\vec\s_1)$};
        \node[rectangle, draw] (0b) [below of =0, yshift = -0.2cm]{$\begin{array}{c} \text{Dense} \\ (\text{Softmax})\end{array}$};
        \node[rectangle, draw](0c)[below of =0b, yshift=-0.6cm]{$\begin{array}{c}\text{Euclidean/}\\ \text{Hyperbolic}\\ \text{RNN/GRU}\\\end{array}$};
        \node[](0e)[left of = 0c, xshift = -1cm]{};
        \node[circle, draw](0d)[below of =0c, yshift = -0.3cm]{$\s_0$};

        \draw[->] (0d) -- node [right]{} (0c);
        \draw[->] (0c) -- node [right, xshift = 0.1cm]{$h_1$} (0b);
        \draw[->] (0b) -- node [right]{} (0);
        \draw[->] (0e) -- node [right,above, midway]{$h_0$} (0c);

        \node[circle, draw] (1) [right of = 0, xshift=2.2cm]{$P(\vec\s_2)$};
        \node[rectangle, draw] (1b) [below of =1, yshift = -0.2cm]{$\begin{array}{c} \text{Dense} \\ (\text{Softmax})\end{array}$};
        \node[rectangle, draw](1c)[below of =1b, yshift=-0.6cm]{$\begin{array}{c}\text{Euclidean/}\\ \text{Hyperbolic}\\ \text{RNN/GRU}\\\end{array}$};
        \node[circle, draw](1d)[below of =1c,yshift = -0.3cm]{$\s_1$};

         \draw[->] (1d) -- node [right]{} (1c);
        \draw[->] (1c) -- node [right]{$h_2$} (1b);
        \draw[->] (1b) -- node [right]{} (1);
        \draw[->] (0c)-- node [right,above, midway]{$h_1$} (1c);

        \node[circle, draw] (3) [right of = 1, xshift = 2.2cm]{$P(\vec \s_3)$};
        \node[rectangle, draw] (3b) [below of =3, yshift = -0.2cm]{$\begin{array}{c} \text{Dense} \\ (\text{Softmax})\end{array}$};
        \node[rectangle, draw](3c)[below of =3b, yshift=-0.6cm]{$\begin{array}{c}\text{Euclidean/}\\ \text{Hyperbolic}\\ \text{RNN/GRU}\\\end{array}$};
        \node[circle, draw](3d)[below of =3c,yshift = -0.3cm]{$\s_2$};

        \draw[->] (3d) -- node [right]{} (3c);
        \draw[->] (3c) -- node [right]{$h_3$} (3b);
        \draw[->] (3b) -- node [right]{} (3);
        \draw[->] (1c)-- node [right, above, midway]{$h_2$} (3c);

        \node[circle, draw] (4)[above of = 1, yshift= 0.6cm] {$\Psi(\vec\sigma)$};
        \draw[->] (1) -- node [right]{} (4);
        \draw[->] (0) -- node [right]{} (4);
        \draw[->] (3) -- node [right]{} (4);
    \end{tikzpicture}
    \caption{Schematic of the process of calculating the  RNN wavefunction $\Psi(\vec\s)= \sqrt{P(\vec\s)}|\vec\s\rangle$ from the probability $P(\vec\s)$ of the sample $\vec\s$. Here $P(\vec\s) = P(\s_1)P(\s_2|\s_1)\ldots P(\s_N|\s_{N-1})$. For a compact representation, $N=3$ in the schematic. The recurrent network in the diagram can be either Euclidean RNN/GRU or Hyperbolic RNN/GRU. Figure adapted from \cite{hld-hypnqs-25}.} \label{rnn-wavefunc-gen}
    \end{figure}
\FloatBarrier
The autoregressive RNN-based real NQS wavefunction, whose schematic is shown in Fig.\ref{rnn-wavefunc-gen}, consists of a layer of RNN/GRU followed by a dense layer of 2 units with the Softmax activation function. The RNN/GRU layer can be Euclidean or hyperbolic, with the latter choice includes both the options of Poincar\'e and Lorentz models of hyperbolic space. The structure of the RNN-based real NQS function is illustrated in Fig.\ref{rnn-wavefunc-gen}. At a step $i$ where $0\leq i\leq N$, the RNN/GRU cell takes in two arguments, one being the previous RNN/GRU hidden state $\vec h_{i-1}$ and the input $\vec\s_{i-1}$ (the one-hot-encoded $(i-1)^{th}$ component of the generated spin sample $\vec\s=(\s_1, \ldots, \s_N)$), to calculate the current RNN/GRU hidden state $\vec h_i$. This is then passed into the Dense layer with the Softmax activation function\footnote{Recall that the Softmax function is defined as \beq \text{Softmax}(v_k) = \frac{\exp(v_k)}{\sum_i\exp(v_i)}\eeq} to obtain an output $\vec y_i$, which is multiplied with the one-hot-encoded sample $\sigma_i$ to calculate the probability $P(\vec\sigma_i)$. The total probability $P(\vec\s)$ for the single spin sample $\vec\s$ is the product of all individual $P(\vec\s_i)$.
\item \textbf{Complex NQS wavefunction }: The complex NQS wavefunction is defined as
\beq
\Psi(\vec\s) = \sum_{\vec \s}\exp(i\phi(\vec\s))\sqrt{P(\vec\s)}|\vec\s\rangle\,.
\eeq
where $P(\vec\s)$ has the same meaning as the real NQS case, while $\phi(\vec\s)$ is the phase of the generated spin sample $\vec\s$.
\begin{figure}[H]
\centering
\begin{tikzpicture}[node distance = 1.3cm, thick]%
        \node[circle, draw] (0) {$P_1$};
        \node[circle, draw] (0a)[right of = 0] {$\phi_1$};
        \node[circle, draw] (0b) [below of =0]{$S$};
        \node[circle, draw] (0b2) [below of =0a]{$SS$};
        \node[rectangle, draw](0c)[below of =0b, xshift =0.5cm, yshift=-0.4cm]{$\begin{array}{c}\text{Euclidean/} \\\text{Hyperbolic}\\\text{(Poincar\'e/Lorentz)}\\\text{RNN/GRU}\\\end{array}$};
        \node[](0e)[left of = 0c, xshift = -1cm]{};
        \node[circle, draw](0d)[below of =0c, yshift = -0.4cm]{$\s_0$};

        \draw[->] (0d) -- node [right]{} (0c);
        \draw[->] (0c) -- node [right]{$h_1$} (0b);
        \draw[->] (0c) -- node [right]{} (0b2);
        \draw[->] (0b) -- node [right]{} (0);
        \draw[->] (0b2) -- node [right]{} (0a);
        \draw[->] (0e) -- node [right,above, midway]{$h_0$} (0c);

        \node[circle, draw] (1) [right of = 0a, xshift=1.8cm]{$P_2$};
        \node[circle, draw] (1a) [right of = 1]{$\phi_2$};
        \node[circle, draw] (1b) [below of =1]{$S$};
        \node[circle, draw] (1b2) [below of =1a]{$SS$};
        \node[rectangle, draw](1c)[right of =0c, xshift=3.4cm]{$\begin{array}{c}\text{Euclidean/}\\ \text{Hyperbolic}\\\text{(Poincar\'e/Lorentz)}\\ \text{RNN/GRU}\\\end{array}$};
        \node[circle, draw](1d)[below of =1c,yshift = -0.4cm]{$\s_1$};

        \draw[->] (1d) -- node [right]{} (1c);
        \draw[->] (1c) -- node [right]{$h_2$} (1b);
        \draw[->] (1b) -- node [right]{} (1);
        \draw[->] (1b2) -- node [right]{} (1a);
         \draw[->](1c) -- node [right]{} (1b2);
        \draw[->] (0c)-- node [right,above, midway]{$h_1$} (1c);

        \node[circle, draw] (3) [right of = 1a, xshift = 2.2cm]{$P_3$};
        \node[circle, draw] (3a) [right of = 3]{$\phi_3$};
        \node[circle, draw] (3b) [below of =3]{$S$};
        \node[circle, draw] (3b2) [below of =3a]{$SS$};
        \node[rectangle, draw](3c)[right of = 1c, xshift=3.5cm]{$\begin{array}{c}\text{Euclidean/}\\ \text{Hyperbolic}\\\text{(Poincar\'e/Lorentz)}\\ \text{RNN/GRU}\\\end{array}$};
        \node[circle, draw](3d)[below of =3c,yshift = -0.4cm]{$\s_2$};

        \draw[->] (3d) -- node [right]{} (3c);
        \draw[->] (3c) -- node [right]{$h_3$} (3b);
        \draw[->] (3b2) -- node [right]{} (3a);
        \draw[->] (3b) -- node [right]{} (3);
        \draw[->](3c) -- node [right]{} (3b2);
        \draw[->] (1c)-- node [right, above, midway]{$h_2$} (3c);

        \node[circle, draw] (4)[above of = 1, xshift = 1.0cm, yshift = 0.8cm] {$\Psi(\vec\sigma)$};
        \draw[->] (1) -- node [right]{} (4);
        \draw[->] (1a) -- node [right]{} (4);
        \draw[->] (0) -- node [right]{} (4);
        \draw[->] (0a) -- node [right]{} (4);
        \draw[->] (3) -- node [right]{} (4);
        \draw[->] (3a) -- node [right]{} (4);

    \end{tikzpicture}
    \caption{Schematic of the calculation of the RNN wavefunction  $\Psi(\vec\s) = \sum_{\vec\s}\exp(i\phi(\vec\s) \sqrt{P(\vec\s)}|\vec\s\rangle$ where the amplitude $P(\vec\s) = P(\s_1)P(\s_2|\s_1)\ldots P(s_N|\s_{N-1})$ and the phase $\phi(\vec\s) = \sum_{i=1}^N \phi(\vec\s_i)$. For ease of illustration, $N$ is chosen to be 3. (S) and (SS), correspondingly, denote the dense layer with the Softmax and Softsign activation function. Figure adapted from \cite{hld-hypnqs-25}. } \label{rnn-wavefunc-gen-c}
    \end{figure}
The structure of the autoregressive RNN-based complex NQS wavefunction is illustrated in Fig.\ref{rnn-wavefunc-gen-c} where $S$ and $SS$ represent the two Dense layers, one with Softmax activation and one with Softsign activation\footnote{The Softsign function is defined as \beq\text{Softsign}(x) = \frac{x}{1+|x|}\in (-1,1)\,.\eeq}. Overall, this structure differs from that of the real RNN-based NQS shown in Fig.\ref{rnn-wavefunc-gen} by the additional Softsign Dense layer  that is used to generate the phase $\phi(\vec\s)$. The computation of the real part $P(\vec\s)$ is the same as described in the real NQS case above, and will not be repeated here. To compute $\phi(\vec\s)$, the hidden state $\vec h_i$ from the RNN/GRU cell is passed into the Softsign Dense layer to generated the output $\vec y^{(2)}_i$, which is then multiplied with $\vec\s_i$ (the one-hot-encoded $i^{th}$ component of the spin sample $\vec\s$) to obtain $\phi_i$. The final phase of the spin configuration $\vec\s=(\s_1, \ldots, \s_N)$ is the sum of the individual $\phi_i$ phases.
\\\\
In this paper, we only work with the complex NQS wavefunction, since the Hamiltonian systems of interest to us are the Heisenberg models. Besides the details described above, we note that the Marshall sign \cite{j1j2-marshall}  was used in all experiments. As such, the wavefunction $\Psi(\vec\s)$ is modified to be
\beq
\Psi_M(\vec\s) = (-1)^{M_A(\vec\s)}\Psi(\vec\s) \label{marshall}
\eeq
where $M_A(\vec\s) = \sum_{k\in A}\s_k$, with $A$ containing the sites of either all even or all odd positions in the lattice. The factor $(-1)^{M_A(\vec\s)}$ is the Marshall sign of the wavefunction, while $\Psi(\vec\s)$ is the complex wavefunction before the application of the Marshall sign. We note that Eq.(\ref{marshall}) is exact only for the case of unfrustrated Heisenberg systems with only $J_1$, representing nearest neighbor interactions, nonvanishing ($J_{i\geq 1} = 0$). In the case $J_{i\geq 1} \neq 0$ where multiple degrees of nearest neighbor interactions are present, Eq.(\ref{marshall}) only hold approximately.
\end{itemize}
\textbf{Further remarks}: Before moving on to the VMC experiments in the specific Heisenberg models, we note the following details that are common to all experiments done in this work.
\bei
\item Due to our limited computing resources, we restricted the sizes of all neural networks to be relatively small, ranging on the orders of a few to tens of thousands parameters only, in order to carry out a large number of experiments. Our aim is not to use very large models to get a very high level of quantitative accuracy in the VMC energy (e.g. up to six digits) but to benchmark the qualitative performances of the hyperbolic RNN/GRU NQS ans\"atze with respect to the Euclidean RNN/GRU NQS ans\"atze, whose capability to attain a very high level of accuracy was already demonstrated in \cite{rnn_20}, \cite{rnn-24}.
\item
During the training phase, the number of spin configurations generated is fixed at 80 for all six types of NQS ans\"atze (Euclidean RNN, Poincar\'e RNN, Lorentz RNN, Euclidean GRU, Poincar\'e GRU, Lorentz GRU) used in this work. For inference, we used $10^4$ spin configuration samples. As explained in \cite{hld-hypnqs-25}, the  choice of the relatively small number of 80 samples\footnote{During the training process, we used two kinds of sampling: with and without the `temperature' parameter $\tau$ (set to 1.05 at the start of the training, decaying to 1 after 500 epochs) which encourages the model to stay in the exploration phase longer (before convergence). When $\tau \neq 1$, the model is `encouraged' to sample more diversely, while when $\tau=1.0$, the model defaults to its natural sampling state. For Poincar\'e GRU/RNN NQS as well as Euclidean RNN/GRU NQS, there were no noticeable differences in the obtained results with or without $\tau$, but for Lorentz RNN/GRU NQS, the trainings with $\tau$ involved seemed to be marginally better than the ones without. Qualitatively this makes sense because Poincar\'e disk is a compact region while Lorentz hyperboloid is more open so in the latter case more explorations might lead to better optimizations while in the former case they might not. In the Euclidean case, qualitatively, the fact that $\tau$ did not lead to a difference is probably due to the flat optimization landscape which does not benefit as much as the open hyperboloid landscape that is harder to navigate. For inference, $\tau$ was not used in the sampling process (since the models had converged). } (during training) was made based on the known fact that RNN-based, autoregressive NQS do not require too many samples to achieve convergence since an autoregressive architecture allows for an exact and independent sampling process by forward-generating spins from conditional distributions (with no Markov chain Monte Carlo sampling required). Furthermore, we used the practice of saving the best model during training, as explained in detail in \cite{hld-hypnqs-25}, which means that the weights and biases of the model under training were saved only if the model showed improvement both in terms of a lower mean energy and a variance being under a specified variance tolerance threshold.

\item For the Euclidean NQS ans\"atze, we used Adam optimizer with a scheduled learning rate decay contingent on the model reaching an energy plateau. For the Poincar\'e hyperbolic NQS ans\"atze, Adam optimizer was used for the Euclidean parameters (consisting of the weight matrices) and Riemannian SGD optimizer \cite{rsgd}, \cite{rsgd-ganea} , \cite{rsgd-ganea-2} was used for the hyperbolic parameters (the biases). For the Lorentz hyperbolic NQS ans\"atze, both the weight matrices and the biases are Euclidean so Adam optimizer was used for all parameters. Similar to the Euclidean case, the same scheduled learning rate decay mechanism was applied to the Poincar\'e and Lorentz neural network training.
We used the learning rate of $5\times 10^{-3}$ for the Euclidean parameters of all networks, while for the hyperbolic parameters (the biases), which are sensitive to the manifold geometry, the best learning rate had to be chosen via a process of trial and error to obtain the most stable training process. As such, the learning rate of the hyperbolic parameters can either be the same as the Euclidean parameters or different, depending on the problem at hand. For all our experiments, the hyperbolic learning rates are typically one of the following four choices: $5\times 10^{-4}$, $1\times 10^{-3}$, $5\times 10^{-3}$, $8\times 10^{-3}$ and these are tested on each hyperbolic NQS to determine the one that yielded the best result. Furthermore, it must be noted that for the Lorentz networks, although the biases are Euclidean, they are transformed via the exponential map to be hyperbolic variables to be added on the manifold, so they can be subjected to a hyperbolic learning rate while the strictly Euclidean weight matrices are subjected to the Euclidean learning rate.

\item As noted in \cite{hld-hypnqs-25},  hyperbolic networks were expected to take much longer to train than Euclidean networks due to their much more complicated mathematical constructions involving manifold-native operations. While this is still the case for the experiments done in this work, the training durations of the Poincar\'e hyperbolic networks have been  significantly improved by a factor of at least 5 times compared to our previous work, thanks to some optimizations carried out in the new \texttt{Pytorch} code which eliminated the redundancies present in the previous \texttt{TensorFlow} codes in the probability calculations of the NQS ans\"atze. In our previous work \cite{hld-hypnqs-25}, we noted that Poincar\'e GRU could take more than 10 hours to train, but with the new code, the training time for Poincar\'e GRU is only around twice that of Euclidean GRU (around 2-3 hours). Poincar\'e RNN and Lorentz RNN are relatively fast to train, with the time taken being comparable to that taken by Euclidean GRU. Lorentz GRU, on the other hand, took much longer to train (up to more than 10 hours) than Poincar\'e GRU.

\item Finally, for both Euclidean and hyperbolic networks, we used gradient clipping during the training process to ensure stable convergence. However, unlike the Euclidean RNN and GRU NQS ans\"atze which almost always delivered a stable behavior during training with just gradient clipping, due to the exponential growth of space in hyperbolic geometry, both Poincar\'e and Lorentz hyperbolic networks are naturally prone to additional numerical instabilities during the training process.
Recall that Poincar\'e model is a bounded disk with an edge of radius $c=1.0$, while Lorentz model is an open hyperboloid embedded in the Minkowski space with one extra spatial dimension. Instabilities in the form of `numerical explosions' can occur in either model, with different underlying causes. For Poincar\'e RNN/GRU, unstable behaviors typically happen when the norm of the hidden state $\vec h_i$ is pushed to the edge of the Poincar\'e disk (at 1.0), reaching the value of 0.9999. For Lorentz RNN/GRU, unstable behaviors typically happened when the spatial (and time) components of the hidden state vector $\vec h_i$ grows exponentially large (like $10^{2}, 10^3$), causing the state to be pushed off the hyperboloid. When these numerical explosions happen, the hyperbolic models can  drift around without settling into a minimum or even get trapped in a local mininum, causing the mean energy to stall at a value much higher than the true ground state. For this reason, to make sure that we obtained stable behaviors, a maximal radius $R_{max}<c$ was chosen for the Poincar\'e NQS, and a maximal spatial norm $L_{max}$ constraint on the spatial components of all hyperbolic vectors was implemented for the Lorentz models. These necessary measures ensured that the hidden state $\vec h_i$ of the  hyperbolic RNN/GRU models was well-behaved, which translated into stable performances of the hyperbolic networks during the training process. The maximal radius $R_{max}$ in Poincar\'e disk and the maximal spatial norm clamp $L_{max}$ in Lorentz hyperboloid are therefore additional hyperparameters for the hyperbolic networks that must be chosen carefully to obtain optimal performances of  these networks. More details on these hyperparameters for each specific Heisenberg model will be discussed in the sections below, as well as in the Appendix, Section \ref{spatial-constr}.
\eni

\section{1D Heiseinberg $J_1J_2$ Model}\label{sec-1dj1j2}
\begin{figure}[!h]
\centering
\includegraphics[width = .7\textwidth]{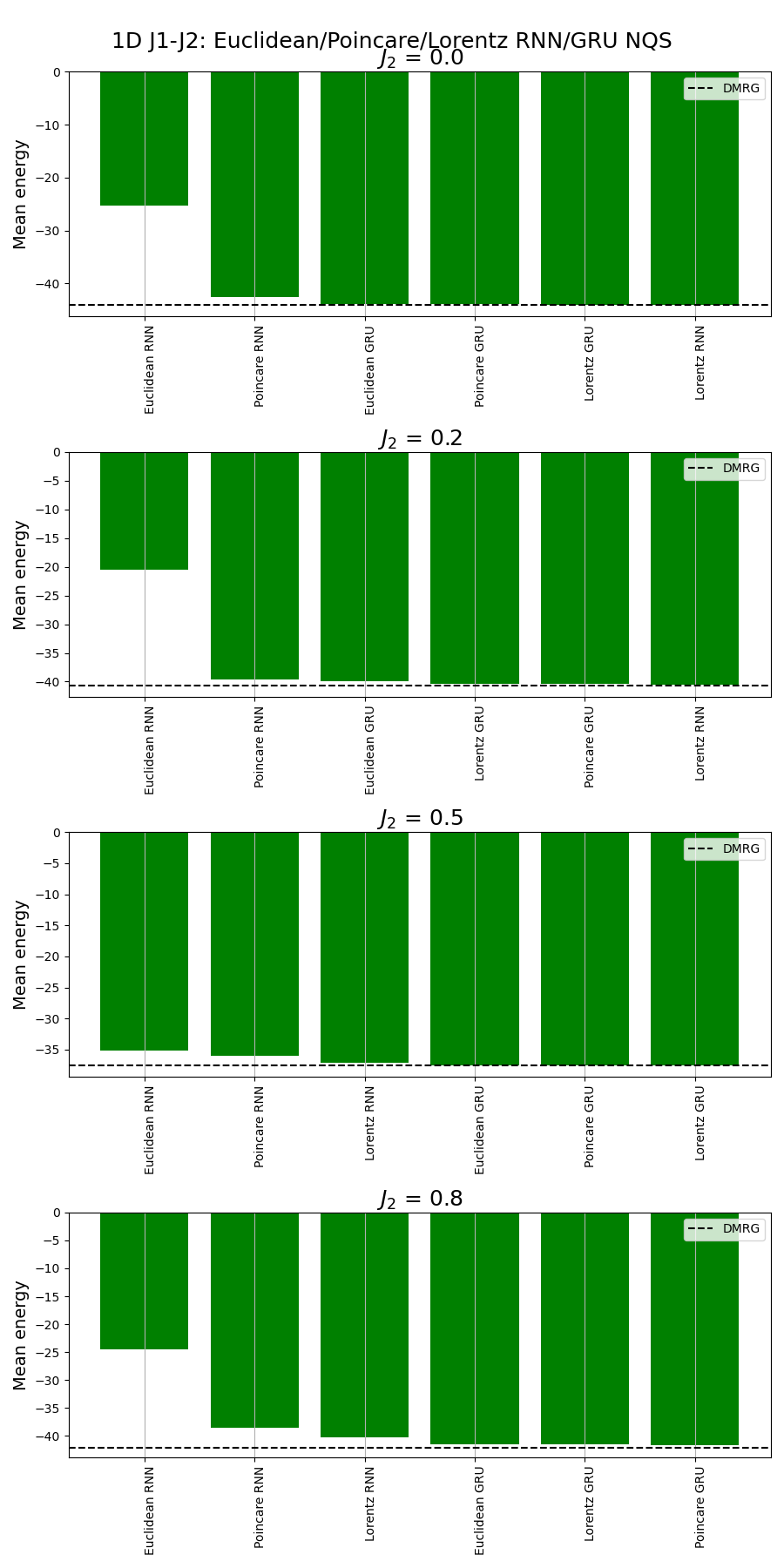}
\caption{Comparisons of the best achievable performances of all six NQS ans\"atze, sorted in an ascending performance order for the  1D Heisenberg $J_1J_2$ model with $J_1 = 1.0$ and different $J_2$ couplings - top to bottom: $J_2 = 0.0, 0.2, 0.5, 0.8$. Note that all RNN variants have 5394 parameters, while all GRU variants have 15614 parameters. }\label{j1j2-all-comparison}
\end{figure}

\begin{figure}[!h]
\centering
\includegraphics[width = .7\textwidth]{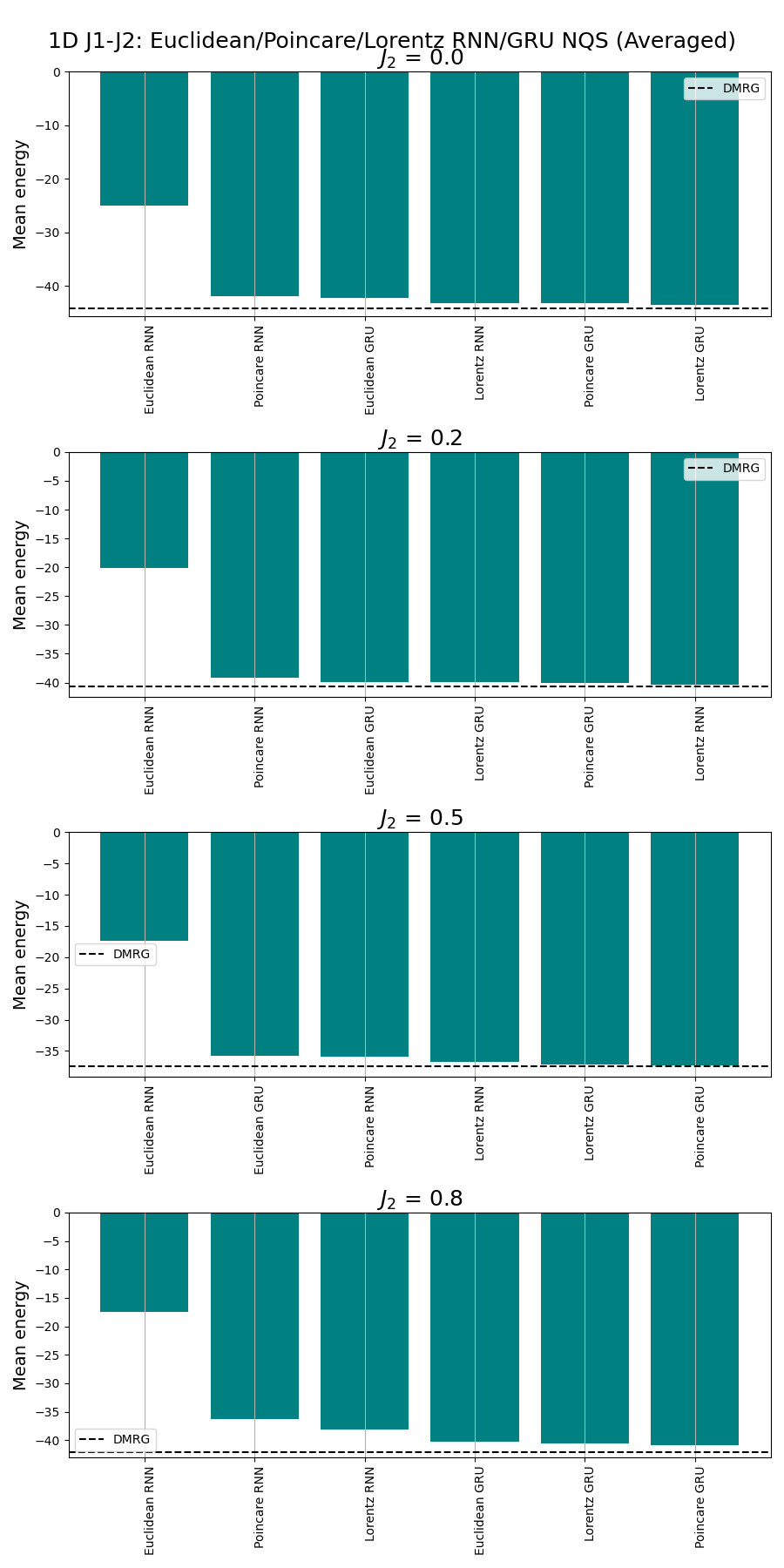}
\caption{Comparisons of the averaged performances of  all six NQS ans\"atze, sorted in an ascending \textit{averaged} performance order for the  1D Heisenberg $J_1J_2$ model with $J_1 = 1.0$ and different $J_2$ couplings - top to bottom: $J_2 = 0.0, 0.2, 0.5, 0.8$. Note that all RNN variants have 5394 parameters, while all GRU variants have 15614 parameters. }\label{j1j2-all-comparison-ave}
\end{figure}

The first Hamiltonian system of interest to us is the frustrated 1D Heisenberg $J_1 J_2$ model
\beq
H_{J1J2} = J_1 \sum_{\langle i,j\rangle} S_i S_j + J_2 \sum_{\langle\langle i, j\rangle\rangle} S_i S_j, 
\eeq
with open boundary condition, where $S_{i}$ denoting the Pauli spin half operators. The Hamiltonian is a sum of two terms: the first takes into account the nearest neighbor interactions pairs $\langle i,j\rangle$, the second takes into account the next nearest neighbor interaction pairs $\langle\langle i,j\rangle\rangle$. In this work, we are interested in the case of antiferromagnetic couplings where both $J_1>0$ and $J_2>0$.  The physics of the 1D $J_1J_2$ model at different couplings $J_2$ is well studied \cite{j1j2-dmrg} and is very briefly summarized below.
\bei
\item At $J_2=0.0$, the system exists in a gapless Luttinger liquid phase with translation symmetry.
\item At $J_2=0.2$, the system still exists in the Luttinger liquid phase, albeit with frustration, with translation symmetry intact. At $J_2 = 0.2412$, the system exhibits a phase transition from a critical Luttinger liquid phase to a spontaneously gapped valence phase, more generally known as the `dimer' phase where adjacent spins form pairs and translation symmetry is broken.
\item At $J_2 = 0.5$, the second neighbor coupling is exactly half as strong as the first neighbor coupling $J_1$, the model reduces to the Majumdar-Ghosh model where the quantum system is exactly solvable and the ground state is a perfectly commensurate dimer phase (where the spins alternate exactly on every other site).
\item At $J_2 =0.8$, the system is highly frustrated as $J_2$ is almost as large as $J_1$, and the ground state is generally known to exist in an incommensurate dimer phase (where a spiral pattern exists alongside the spin alignment).
\eni
 To study the performances of Poincar\'e and Lorentz hyperbolic RNN/GRU NQS ans\"atze versus the Euclidean RNN/GRU NQS ans\"atze, we fixed the number of spins in the system to be $N=100$ (exactly two times larger than the $J_1J_2$ system studied in \cite{hld-hypnqs-25}), the coupling $J_1=1.0$ and varied $J_2=0.0,0.2,0.5,0.8$ in the same fashion as in our earlier work \cite{hld-hypnqs-25}. The exact energy of the four Heisenberg models were calculated with DMRG (Density Matrix Renormalization Group) method.
\\\\
We list the six types of NQS wavefunctions that were used to run VMC experiments for this system in Table \ref{j1j2-ansatz}. As already noted in \cite{hld-hypnqs-25}, the aim of our work is to establish the performances of the new hyperbolic RNN/GRU NQS ans\"atze relative to the well-established Euclidean RNN/GRU NQS ans\"atze (\cite{rnn_20}, \cite{rnn-24}), so the inclusion of the Euclidean RNN/GRU NQS ans\"atze in our experiments served purely as benchmarks for the new hyperbolic NQS.
For the $J_1J_2$ experiment settings, all RNN/GRU had the same $\vec h_i$ hidden dimension size of 70 units, meaning that the RNN variants had 2.895 times fewer parameters than the GRU variants (where each GRU/RNN variant contains three different types of neural networks: Euclidean, Poincar\'e and Lorentz). This choice was motivated by the fact that in designing our experiments, the size of the hidden dimension $\vec h$ of the RNN/GRU mattered to us more than the total number of parameters. The results listed in Table \ref{j1j2-RNN-GRU} correspond to the best achievable results for each of the six NQS ans\"atze chosen from multiple VMC runs while those listed in Table \ref{j1j2-RNN-GRU-ave} correspond to the averaged results for each NQS ansatz taken from all VMC runs, where a single VMC run is defined by a distinct random seed, where each seed corresponds to a different initialization configuration before training starts. The full results corresponding to all individual random seeds are listed in Section \ref{full-res-j1j2} for all six types of NQS.
\\\\
 For this set of experiments, we fixed the number of training epochs to 1000, with an early stopping mechanism built in to terminate the training early in case no improvement was registered for a period of 200 epochs. As such, the number of training epochs varied for each NQS. As mentioned above, we used a variable learning rate scheme \texttt{ReduceLROnPlateau} which automatically decays the learning rate (for both Euclidean and hyperbolic parameters) by a factor of 2.0 when no improvements are detected within 40 epochs\footnote{Besides the main difference being \texttt{Pytorch} rather than \texttt{TensorFlow} in this work compared to \cite{hld-hypnqs-25}, the use of \texttt{ReduceLROnPlateau} differed from the learning rate decay mechanism used in \cite{hld-hypnqs-25} in which the learning rate of Euclidean parameters was subjected to a decay factor of 0.9 every 100 epochs while the hyperbolic learning rate was kept fixed, with both initial learning rates set to be $10^{-2}$.}. The initial learning rate for Euclidean parameters of both Euclidean and hyperbolic networks was fixed at $5\times 10^{-3}$, while the initial learning rate for hyperbolic parameters was chosen on a case-by-case and trial-and-error basis. Most often, the initial hyperbolic learning rate was fixed at either $5\times 10^{-3}$ (same as Euclidean parameters) or $8\times 10^{-3}$.
\\\\
Also as mentioned previously, Euclidean NQS were straightforward to train and they almost always delivered stable performances, largely thanks to the fact that besides choosing the number of RNN/GRU hidden dimensions and the initial learning rate for the Euclidean parameters, there is no other hyperparameter that required special attention. On the other hand, hyperbolic networks have more hyperparameters in the form of the hyperbolic learning rate as well as an additional hyperparameter - the maximal radius $R_{max}$ for the Poincar\'e networks or the maximal spatial norm $L_{max}$ for the Lorentz networks, which we will collectively refer to as the spatial constraint hyperparameter - that must be chosen very carefully to ensure the best performances that hyperbolic network can deliver. Unsuitable choices of this spatial constraint hyperparameter lead to poor and unstable performances by hyperbolic networks (see the discussion in Section \ref{spatial-constr} in the Appendix).
The specific choices (obtained through a rather laborious and tedious process of trial-and-error) of $R_{max}$ and $L_{max}$ for the hyperbolic networks are listed in Table \ref{j1j2-hyperparams-poincare} and \ref{j1j2-hyperparams-lorentz}, respectively (Section \ref{spatial-constr}).
\begin{table}[!ht]
\centering
\begin{tabular}{clccc}
\hline\hline
& Ansatz  & Hidden units & Parameters &
\\\hline\hline
& Euclidean RNN& 70  &  5394 &  \\
&Lorentz RNN  & 70   &  5394 &  \\
&Poincar\'e RNN& 70 & 5394& \\
&Euclidean GRU &70  & 15614  &  \\
&Lorentz GRU  &70   &  15614 &  \\
& Poincar\'e GRU& 70 & 15614&
\\ \hline
\hline
\end{tabular}
\caption{The six types of NQS wavefunctions used for approximating the ground state energies of the 1D Heisenberg $J_1J_2$ model. These NQS were used in all cases of different $J_2$ couplings studied.} \label{j1j2-ansatz}
\end{table}
\FloatBarrier
From Table \ref{j1j2-RNN-GRU} and Table \ref{j1j2-RNN-GRU-ave}, as well as Figs. \ref{j1j2-all-comparison}, \ref{j1j2-all-comparison-ave}, the following observations are made. 
\begin{itemize}
     \item \textit{Architecture-wise}, in terms of the best achievable mean energy (see Table \ref{j1j2-RNN-GRU}), within an architecture variant (either RNN or GRU), for all $J_2$ values, hyperbolic NQS ans\"atze always outperformed their Euclidean counterparts, i.e. Poincar\'e RNN and Lorentz RNN always outperformed Euclidean RNN, and Poincar\'e GRU and Lorentz GRU always outperformed Euclidean GRU. This is the generalization of the $J_1J_2$ results reported in \cite{hld-hypnqs-25} in which we only constructed and introduced a single form of NQS hyperbolic network - the Poincar\'e hyperbolic GRU, which was benchmarked agains the Euclidean GRU for $J_1J_2$ Heisenberg system with 50 spins. More specifically, if we look at each architecture separately:
     \begin{enumerate}
    \item \textit{NQS ans\"atze with RNN architecture}: Among the three RNN variants (with 5394 parameters),  for all $J_2$ couplings, Lorentz RNN is always the best performing NQS, followed by Poincar\'e RNN and Euclidean RNN. In particular, Euclidean RNN very often settles into a local minimum that is much higher than the true ground state and lags far behind both Poincar\'e and Lorentz RNN in terms of performance.
   Between the two hyperbolic RNNs, Poincar\'e RNN never even attains the level of performance comparable to Lorentz RNN, despite the supposed one-to-one mapping between  Poincar\'e RNN and Lorentz RNN. At the moment, it is not clear to us whether this underperformance by Poincar\'e RNN can be more attributable to the lack of a comprehensive $R_{max}$ tuning or more to the inherent nature of the Poincar\'e RNN's construction in the Poincar\'e disk.
    \item \textit{NQS ans\"atze with GRU architecture}: Among the three GRU variants (with 15614 parameters), depending on the specific $J_2$ value, either Lorentz GRU or Poincar\'e GRU is the best performing NQS. In particular, at $J_2= 0.0, 0.5$, Lorentz GRU is the best performing variant, while for $J_2= 0.2, 0.8$, Poincar\'e GRU ranks first. Euclidean GRU always ranks third among the three GRU variants.
    The gaps between the performances of Euclidean GRU and its hyperbolic counterparts are very small, unlike the gaps between Euclidean RNN and its hyperbolic versions. For all $J_2$ couplings, while Lorentz RNN clearly outperformed Poincar\'e RNN, the opposite is true for Poincar\'e GRU and Lorentz GRU, with each occupying the top spot two out of four times, making it an even split between Lorentz GRU and Poincar\'e GRU for the $J_1J_2$ system.
\end{enumerate}
   \item \textit{Across architectures}: When all six NQS ans\"atze are compared, as shown in Figs.\ref{j1j2-all-comparison} in terms of the best achievable mean energy, a very interesting picture emerged. While Euclidean RNN is consistently the worst performing variant, with Poincar\'e RNN consistently the second worst, the top four spots are held by Euclidean GRU, Poincar\'e GRU, Lorentz GRU and Lorentz RNN, in different orders depending on the specific value of $J_2$.
    \begin{itemize}
        \item $J_2=0.0$: Lorentz RNN is the top performer, with Lorentz GRU coming in second, followed by Poincar\'e GRU (third) and Euclidean GRU (fourth).
        \item $J_2 = 0.2$: Lorentz RNN again ranks first, followed by Poincar\'e GRU (second), Lorentz GRU (third),  and Euclidean GRU (fourth).
        \item $J_2 = 0.5$: Lorentz GRU is the top performer, followed by Poincar\'e GRU (second), Euclidean GRU  (third), Lorentz RNN  (fourth).
        \item $J_2 = 0.8$: Poincar\'e GRU is the top performer, followed by Lorentz GRU (second), Euclidean GRU (third),  Lorentz RNN (fourth).
    \end{itemize}
 Among the four hyperbolic NQS ans\"atze, Poincar\'e RNN seems to be the least performant variant, as it underperformed Euclidean GRU, Lorentz RNN, Poincar\'e GRU and Lorentz GRU in all four experiments. While it was expected that the hyperbolic GRU ans\"atze would outperform Poincar\'e RNN and Lorentz RNN thanks to their sophisticated GRU gating mechanism and three times more parameters, Lorentz RNN defied expectations to deliver a surprising overall performance.  On two occasions when $J_2=0.0, 0.2$, Lorentz RNN was able to surpass all GRU variants to be the top performer, despite having almost three times fewer parameters. This might point to the possibility that for the $J_1J_2$ model at these low $J_2$ couplings (corresponding to a low level of frustration in the system), the simpler hyperbolic RNN architecture is superior (in providing a ground state approximation to the Heisenberg model) to the more complex GRU gating mechanism, both in Euclidean and hyperbolic spaces. For $J_2=0.5, 0.8$ where hyperbolic GRU variants (Poincar\'e and Lorentz) ranked in top two, Lorentz RNN underperformed all GRU variants including Euclidean GRU. This might be indicative of the possibility that for these highly frustrated cases where the ground state is in a dimerized phase, the pure hyperbolic geometry offered by Lorentz RNN was insufficient to surpass the sophisticated gating mechanism offered by Euclidean GRU and the combination of hyperbolic geometry and GRU gating mechanism embodied in both the Poincar\'e and Lorentz GRU.
\\\\
 When the performances are ranked in terms of the averaged mean energy for each NQS ansatz obtained from all the VMC runs taken together, as shown in Fig.\ref{j1j2-all-comparison-ave}, the emerging picture is slightly different but the major qualitative features similar to the `best achievable' case above remain. In particular, Euclidean RNN is still the worst performing variant for all $J_2$ values, with the rest of the NQS ranking in descending order as follows.
     \begin{itemize}
        \item $J_2=0.0$: Lorentz GRU is the top performer, with Poincar\'e GRU  in the second spot, followed by Lorentz RNN (third) Euclidean GRU (fourth) and Poincar\'e RNN (fifth).
        \item $J_2 = 0.2$: Lorentz RNN ranks first, followed by Poincar\'e GRU (second), Lorentz GRU (third),  and Euclidean GRU (fourth) and Poincar\'e RNN (fifth). This is the same as the `best achievable' case above.
        \item $J_2 = 0.5$: Poincar\'e  GRU ranks first, followed by Lorentz GRU (second), Lorentz RNN (third), Poincar\'e RNN (fourth) and Euclidean GRU (fifth). This is the only time when Poincar\'e RNN ranks above Euclidean GRU in the $J_1J_2$ experiments.
        \item $J_2 = 0.8$: Poincar\'e GRU ranks first, followed by Lorentz GRU (second), Euclidean GRU (third),  Lorentz RNN (fourth) and Poincar\'e RNN (fifth).
    \end{itemize}
From multiple VMC runs (see the full results for all NQS ans\"atze with different random seeds listed in Section \ref{full-res-j1j2} in the Appendix), we notice that for some of the random seeds, Poincar\'e RNN (see Table \ref{full-j1j2-vmc-prnn}) and Lorentz RNN/GRU (Tables \ref{full-j1j2-vmc-lrnn} and \ref{full-j1j2-vmc-lgru}) failed to converge (i.e. the mean energy either settled into an artificially high value far away from the true ground state, e.g. like in the positive value range, or some sort of kinks or steps  occurring in the training). Out of the four hyperbolic NQS, Poincar\'e GRU showed the most stable performance with no non-convergence issues (see Table \ref{full-j1j2-vmc-pgru}), with obtained energy values differing little across different seeds. Poincar\'e RNN was most stable  when $J_2=0.0, 0.2$ with non-convergence issues for some of the seeds surfacing when $J_2=0.5, 0.8$. Lorentz RNN was most stable when $J_2=0.2, 0.5$ with less stability shown when $J_2=0.0, 0.8$. Lorentz GRU was most stable when $J_2=0.0, 0.2$ and less stable when $J_2=0.5, 0.8$. However, the values obtained by all hyperbolic NQS when they did converge were consistently within a narrow range of one another without significant deviations.  The instability of the hyperbolic NQS training is directly attributable to the fact that the Poincar\'e and Lorentz hyperbolic optimization landscapes, with their negative curvatures determined by the spatial constraints hyperparameters $R_{max}$ or $L_{max}$, are much more `'rugged' compared to the flat Euclidean optimization landscape. This is a challenge well known and well documented from NLP works dealing with hyperbolic neural networks \cite{ganea-1805}, \cite{hyprnn-lorentz-2105}, \cite{1805-hyp-attn}, \cite{2006-hypcnn}.
\\\\
 While the challenge associated with hyperbolic optimization leads to the expectations that Euclidean variants should be much more stable in their trainings compared to their hyperbolic counterparts, interestingly, in our experiments, even the Euclidean variants showed wild fluctuations in terms of the obtained mean energies across different runs for  some values of $J_2$. For example, at $J_2=0.5$ (where the energy is -37.5) at the Majumdar-Ghosh point, both Euclidean RNN NQS and Euclidean GRU NQS (see Tables \ref{full-j1j2-vmc-ernn} and \ref{full-j1j2-vmc-egru}) obtained inconsistent results. In four of five runs, Euclidean RNN NQS converged at the false minimum of around $-12$ to $-13$, with single run where it could achieve a much closer value of $-35.06$, which was its best achievable energy recorded in Table \ref{j1j2-RNN-GRU}.  In two out of five runs, Euclidean GRU NQS converged at the false minimum of around -36 and -30, while on the other three runs, it could reach below -37.  In another example, when $J_2=0.8$ with $E=-42.07$, Euclidean RNN NQS converged at values ranging from  $-5.99$ to $-24.56$. These observations highlight the fact that even for the stable and established Euclidean RNN/GRU NQS, statistical fluctuations in convergence characteristics do occur across multiple random seeds, so the training behaviors of the hyperbolic NQS observed in this work are within normal expectations, given the more challenging nature of the hyperbolic optimization process.
\end{itemize}

\begin{table}[!ht]
\centering
\begin{tabular}{lcc ccc}
\hline\hline
Ansatz &  $J_2 = 0.0$  & $J_2 = 0.2$ & $J_2 = 0.5$ & $J_2 = 0.8$ &
\\\hline\hline
Euclidean RNN  &  -25.2492  &  -20.4812  &  -35.0857  &  -24.5784 &\\
& 0.0047 & 0.0018  & 0.0229 & 0.0436  &\\\hline
Poincar\'e RNN & -42.5914  &  -39.6020 &   -35.9253 &   -38.4681 &\\
& 0.0141 & 0.0100 &  0.0078  &  0.0184 \\ \hline
Lorentz RNN & \textbf{-44.0144} &   \textbf{-40.5609}   & -37.1428  &  -40.1873 &\\
& 0.0049 &  0.0047 & 0.0046 & 0.0154 &\\ \hline 
Euclidean GRU &  -43.8503 &   -39.9616&    -37.4680 &   -41.5246&\\
& 0.0078 &  0.0050 &  0.0024 &  0.0104 &\\ \hline
Poincar\'e GRU   & -43.8982  &  -40.4190   & -37.4861 &   \textbf{-41.6771}&\\
& 0.0079  & 0.0053 &  0.0018 &  0.0044 &\\\hline
Lorentz GRU & -43.9882   & -40.3153  &  \textbf{-37.4890}  &  -41.5325&\\
& 0.0038 & 0.0086 &  0.0016 &  0.0028& 
\\\hline
DMRG& -44.1277 & -40.7388 &   -37.5000 &   -42.0701 &\\
\hline
\end{tabular}{}
\caption{The best achievable VMC results, chosen from different VMC runs corresponding to different random seeds, of 1D $J_1J_2$ Heisenberg models, where $J_1 = 1.0$ and $J_2 = 0.0, 0.2, 0.5, 0.8$, for all six NQS ans\"atze. For each ansatz, the mean energy is recorded in the first line, followed by the standard error directly below in the second line. All GRU variants have the same number of 15614  parameters. The number of samples used for inference is 10000. The best results are noted in bold.} \label{j1j2-RNN-GRU}
\end{table}
\FloatBarrier

\begin{table}[!ht]
\centering
\begin{tabular}{lcc ccc}
\hline\hline
Ansatz &  $J_2 = 0.0$  & $J_2 = 0.2$ & $J_2 = 0.5$ & $J_2 = 0.8$ &
\\\hline\hline
Euclidean RNN  & -25.0599   & -20.1880 &  -17.3729  &  -17.4880 &\\
& 0.0048 & 0.0043 & 0.0118 & 0.0293 &\\\hline
Poincar\'e RNN &-41.9298  &  -39.2267  &  -35.9034  &  -36.2943 &\\
&0.0188 &  0.0140 & 0.0083 & 0.0317&\\\hline
Lorentz RNN  &-43.1626  &  \textbf{-40.3914}   & -36.7115   & -38.1589 &\\
& 0.0114 & 0.0067 & 0.0129  &0.0133&\\\hline

Euclidean GRU  & -42.1911  &  -39.9140   & -35.7951  &  -40.1973 &\\
&0.0085 & 0.0088 & 0.0105 & 0.0106&\\\hline
Poincar\'e GRU  &  -43.1709   & -40.0353  &  \textbf{-37.2695}  &  \textbf{-40.8949} &\\
& 0.0120 & 0.0068  &0.0033&  0.0115&\\\hline
Lorentz GRU &\textbf{-43.4955}   & -39.9257 &   -37.1197   & -40.5030 &\\
& 0.0063 & 0.0062 & 0.0021 & 0.0134&
\\\hline
DMRG& -44.1277 & -40.7388 &   -37.5000 &   -42.0701 &\\
\hline
\end{tabular}{}
\caption{The averaged VMC results of 1D $J_1J_2$ Heisenberg models from all different runs corresponding to different random seeds, where $J_1 = 1.0$ and $J_2 = 0.0, 0.2, 0.5, 0.8$,  with all six NQS ans\"atze. For each ansatz, the mean energy is recorded in the first line, followed by the standard error directly below in the second line. All GRU variants have the same number of 15614  parameters. The number of samples used for inference is 10000. The best results are noted in bold.} \label{j1j2-RNN-GRU-ave}
\end{table}
\FloatBarrier
\FloatBarrier
\clearpage
\FloatBarrier
\clearpage
\section{1D Heiseinberg $J_1J_2J_3$ Model}\label{sec-1dj1j2j3}
\begin{figure}[!h]
\centering
\includegraphics[width = .7\textwidth]{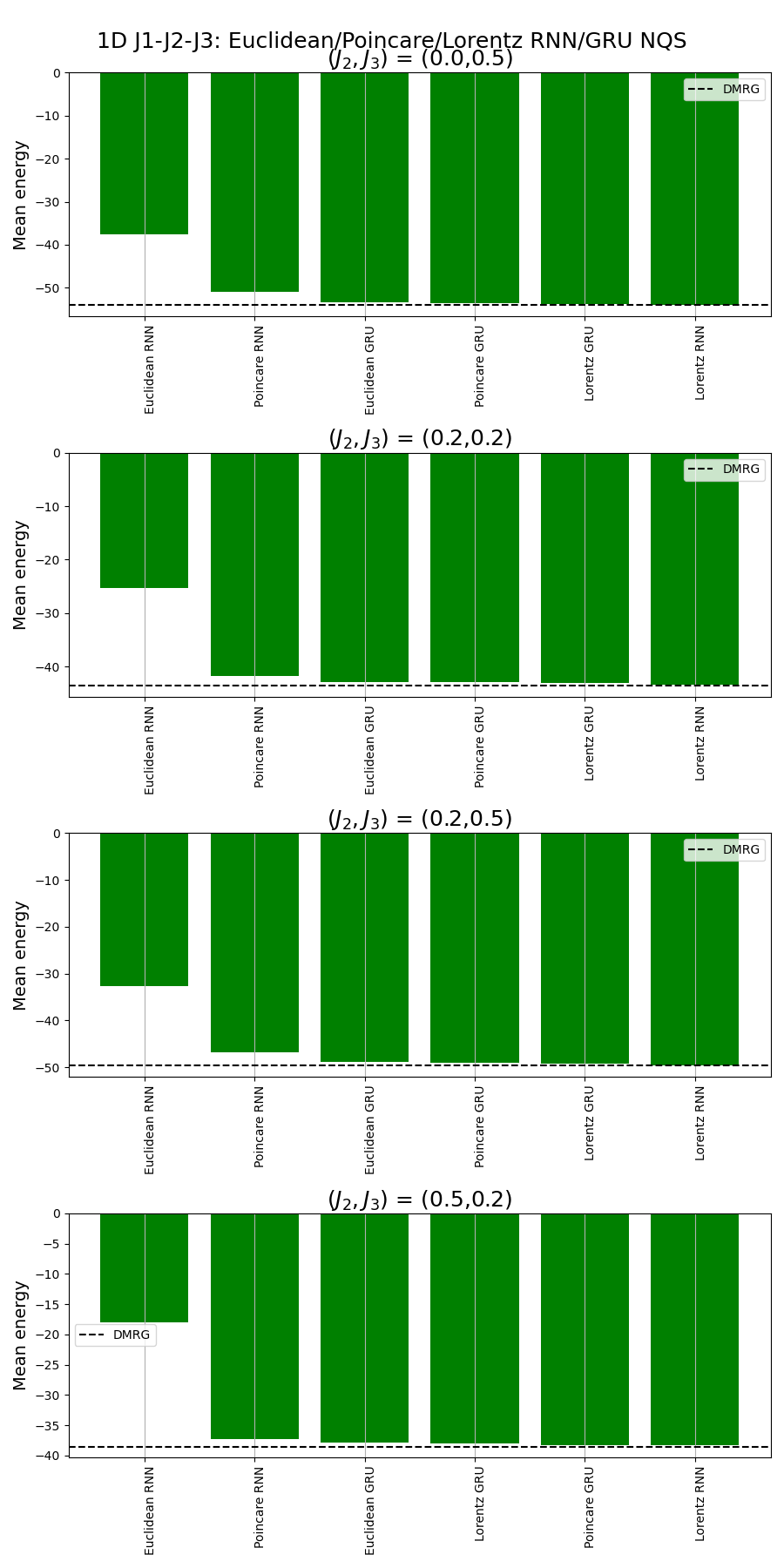}
\caption{Comparisons of the best achievable performances of all NQS ans\"atze, sorted in an ascending performance order for the  1D Heisenberg $J_1J_2J_3$ model with $J_1 = 1.0$ and different $(J_2, J_3)$ couplings - top to bottom: $(J_2, J_3) = (0.0,0.5)$, $(0.2,0.2)$, $(0.2, 0.5)$, $(0.5,0.2)$. Note that the RNN variants have 6964 parameters while the GRU variants have 15614 parameters.}\label{j1j2j3-all-comparison}
\end{figure}

\begin{figure}[!h]
\centering
\includegraphics[width = .7\textwidth]{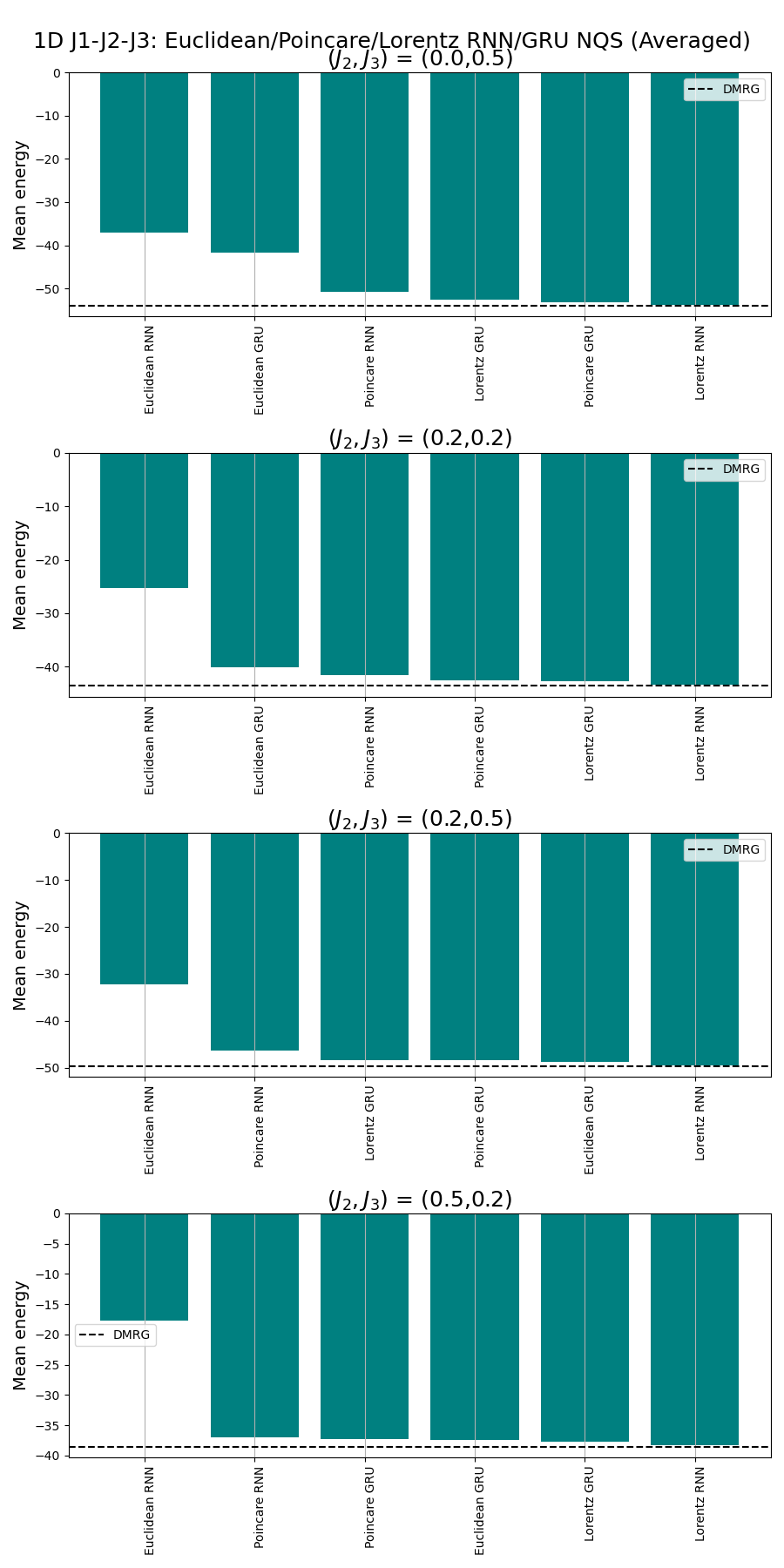}
\caption{Comparisons of the averaged performances of all NQS ans\"atze, sorted in an ascending performance order for the  1D Heisenberg $J_1J_2J_3$ model with $J_1 = 1.0$ and different $(J_2, J_3)$ couplings - top to bottom: $(J_2, J_3) = (0.0,0.5)$, $(0.2,0.2)$, $(0.2, 0.5)$, $(0.5,0.2)$. Note that the RNN variants have 6964 parameters while the GRU variants have 15614 parameters.}\label{j1j2j3-all-comparison-ave}
\end{figure}

The second Hamiltonian system of interest to us is the frustrated 1D Heisenberg $J_1 J_2J_3$ model
\beq
H_{J1J2J3} = J_1 \sum_{\langle i,j\rangle} S_i S_j + J_2 \sum_{\langle\langle i, j\rangle\rangle} S_i S_j
+ J_3 \sum_{\langle\langle\langle i, j\rangle\rangle\rangle} S_i S_j, \eeq
with open boundary condition, where $S_{i}$ denoting the Pauli spin half operators. The Hamiltonian is a sum of three terms: the first takes into account the nearest neighbor interactions pairs $\langle i,j\rangle$, the second takes into account the second nearest neighbor interaction pairs $\langle\langle i,j\rangle\rangle$, and the third takes into account the third nearest neighbor interactions. In this work, we are interested in the $J_1J_2J_3$ model with all antiferromagnetic couplings, i.e. $J_n>0$ for $n=1,2,3$, which means that the system is highly frustrated.
The physics of the 1D $J_1J_2J_3$ model is much less studied than that of the 1D $J_1J_2$ model in the literature. Perhaps the most recent work\footnote{Two dimensional $J_1J_2J_3$ models are more often studied than one-dimensional $J_1J_2J_3$ models. } on the 1D $J_1J_2J_3$ model is \cite{10-1d-j1j2j3}, in which the authors studied the $J_1J_2J_3$ model with a ferromagnetic coupling $J_1$ and two antiferromagnetic couplings $J_2, J_3$ to study the phase transition of the system.
\\\\
In this work, our interest lies in approximating the ground states of the $J_1J_2J_3$ Heisenberg models using the new hyperbolic NQS ans\"atze.
 Similar to the $J_1J_2$ case above, and in the same manner to what was studied in our earlier work \cite{hld-hypnqs-25}, we fixed $J_1=1.0$ and varied the couplings  $(J_2$, $J_3) = (0.0, 0.5)$, (0.2,0.2), (0.2, 0.5) and (0.5, 0.2) to obtain different physical regimes under which the different NQS wavefunctions are used to approximate the true ground state energies. In this work, however, the number of spins is fixed to be $N=100$ (more than three times larger than the case studied in \cite{hld-hypnqs-25}).
 While the exact physics of the 1D $J_1J_2J_3$ in these coupling regimes requires more detailed studies which are beyond the scope of this work whose main aims are to establish the new types of hyperbolic RNN-based NQS ans\"atze, our choices of the $(J_2,J_3)$ couplings here and in \cite{hld-hypnqs-25} were made with the $J_1J_2$ model in mind, namely we added either $J_3=0.2$ or $J_3= 0.5$ to the previously chosen values of  $J_2=0.0,0.2,0.5$ in the $J_1J_2$ model.
 \bei
 \item When $J_3=0.2$, $J_3\leq J_2$ as in (0.2,0.2) and (0.5,0.2): it can be assumed that the physics of the $J_1J_2J_3$ models are reasonably close to that of the $J_1J_2$ model, based on the fact that the ground state energies are these two Heisenberg models are not too far apart, i.e. $E_{J_2=0.5} = -37.5000$ while $E_{(0.5, 0.2)} = -38.54733$, $E_{(0.2,0.2)} = -43.5860$, while $E_{J_2=0.2} = -40.7388$.
 \item  When $J_3=0.5$, $J_3>J_2$ as in (0.0,0.5), (0.2,0.5): the third neighbor interaction is dominant, and the physics of the two models are very different, with their corresponding ground state energies being very far apart, i.e. $E_{J_2=0.0} = -44.1277$, while $E_{(0.0,0.5)} = -53.9914$, $E_{J_2=0.2} = -40.7388$, while $E_{(0.2,0.5)} = -49.6287$.
\eni
We list in Table \ref{j1j2j3-ansatz} the six types of RNN/GRU-based NQS ans\"atze that are used for this system. For the $J_1J_2J_3$ experiment settings, all RNN variants had 80 units in the hidden dimension $\vec h_i$, while the GRU variants had 70 units, meaning that the former had 2.242 times fewer parameters than the latter.
All the details regarding the training of these NQS as well as the choices of the hyperparameters are identical to the $J_1J_2$ case (except for the fact that the number of epochs were fixed to 1200 this time) and will not be repeated here. In Table \ref{j1j2j3-RNN-GRU}, we list the best achievable results for each of the six NQS ans\"atze chosen from multiple VMC runs while in Table \ref{j1j2j3-RNN-GRU-ave}, we list the averaged results for each NQS ansatz taken from all VMC runs. The full results corresponding to all individual random seeds are listed in Section \ref{full-res-j1j2j3} for all six types of NQS.
The choices of the spatial constraint hyperparameters $R_{max}$ and $L_{max}$ for the Poincar\'e and Lorentz hyperbolic neural networks are included in Table \ref{j1j2j3-hyperparams-poincare} and Table \ref{j1j2j3-hyperparams-lorentz} in the Appendix.
\begin{table}[!ht]
\centering
\begin{tabular}{clccc}
\hline\hline
& Ansatz  &  Hidden units & Parameters &
\\\hline\hline
& Euclidean RNN& 80  & 6964  &  \\
&Lorentz RNN    & 80 & 6964 &  \\
&Poincar\'e RNN& 80 & 6964& \\
&Euclidean GRU & 70 & 15614  &  \\
&Lorentz GRU  & 70   &  15614 &  \\
& Poincar\'e GRU & 70& 15614&
\\ \hline
\hline
\end{tabular}
\caption{The six types of NQS wavefunctions used for approximating the ground state energies of the 1D $J_1J_2J_3$ model. These NQS were used in all cases of different ($J_2,J_3$) couplings studied.} \label{j1j2j3-ansatz}
\end{table}
\FloatBarrier
From Table \ref{j1j2j3-RNN-GRU} and Table \ref{j1j2j3-RNN-GRU-ave}, as well as Figs.\ref{j1j2j3-all-comparison}, \ref{j1j2j3-all-comparison-ave}, we can make the following observations regarding the performance ranking of each NQS. Many details are identical to the $J_1J_2$ case studied in the previous section, but there are also some important differences especially when it comes down to comparing the hyperbolic NQS ans\"atze corresponding to different hyperbolic sub-geometries.
\begin{itemize}
    \item \textit{Architecture-wise}, within the same architecture variant (either RNN or GRU), in terms of the best achievable mean energy (see Table \ref{j1j2j3-RNN-GRU} and Fig.\ref{j1j2j3-all-comparison}), hyperbolic NQS (either Lorentz or Poincar\'e) always outperformed Euclidean NQS. This result is the generalization of the $J_1J_2J_3$ results reported in \cite{hld-hypnqs-25} where we established that Poincar\'e GRU almost always outperformed Euclidean GRU for $J_1J_2J_3$ systems with 30 spins. In this more general results, with a larger physical system of 100 spins, we broadened the class of hyperbolic NQS to include not only Poincar\'e GRU but also Lorentz GRU as well as  Poincar\'e and Lorentz RNN. More specifically, if we look at each architecture separately:
    \begin{enumerate}
    \item \textit{NQS ans\"atze with RNN architecture}: Among the RNN variants with  6964 parameters, for all values of $(J_2, J_3$) couplings, Lorentz RNN is always the top performer, followed by Poincar\'e RNN. Euclidean RNN trailed them both by a very large margin. This is identical to what was observed in the 1D $J_1J_2$ case.
    \item \textit{NQS ans\"atze with GRU architecture}: Among the GRU variants with 15614 parameters, on three out of four occasions corresponding to the cases of $(J_2,J_3) = (0.0, 0.5)$, (0.2,0.2), (0.2,0.5), Lorentz GRU is the top performer, followed by Poincar\'e GRU. For the last case of $(J_2, J_3) = (0.5, 0.2)$, Poincar\'e outperformed Lorentz GRU. In all four cases, Euclidean GRU was outperformed by its two hyperbolic versions, which is the same as what was observed in the 1D $J_1J_2$ case. The difference in this case, compared to the $J_1J_2$ case above, is the fact that it is no longer an even split between Lorentz GRU and Poincar\'e GRU but it is Lorentz GRU that outperformed Poincar\'e GRU three out of four times.
    \end{enumerate}
      \item \textit{Across architectures}: When all six NQS ans\"atze are compared (see Fig.\ref{j1j2j3-all-comparison}) in terms of the best achievable mean energy that each ansatz could reach, the emerging picture in ascending performance order is as follows.
    \begin{enumerate}
        \item $(J_2, J_3) = (0.0, 0.5)$:  Euclidean RNN (sixth),  Poincar\'e RNN  (fifth), Euclidean GRU  (fourth), Poincar\'e GRU (third), Lorentz GRU (second), Lorentz RNN  (first). 
        \item $(J_2, J_3) = (0.2, 0.2)$: Exact same order as $(J_2, J_3) = (0.0, 0.5)$. 
        \item $(J_2, J_3) = (0.2, 0.5)$: Exact same order as $(J_2, J_3) = (0.0, 0.5)$ and $(J_2, J_3) = (0.2, 0.2)$. 
        \item $(J_2, J_3) = (0.5, 0.2)$:  Euclidean RNN (sixth), Poincar\'e RNN (fifth), Euclidean GRU (fourth), Lorentz GRU (third), Poincar\'e GRU (second),  Lorentz RNN (first).
    \end{enumerate}
    Among the four hyperbolic NQS ans\"atze with different architectures,  conforming to expectations, Poincar\'e GRU and Lorentz GRU outperformed Poincar\'e RNN in all four experiments due to the two former having the more complex GRU gating mechanism and nearly 2.25 times more parameters than the latter (15614 versus 6964).
    Again defying expections for all values of $(J_2,J_3)$ couplings,  Lorentz RNN showed a remarkable performance against all GRU variants, despite having around 2.25 times fewer parameters. It consistently ranked first  placing ahead of all GRU variants including Euclidean GRU,  Poincar\'e GRU and Lorentz GRU all four out of four times.  Similar to the $J_1J_2$ case, this result again demonstrates the possibility that the pure hyperbolic geometry underlying the Lorentz RNN can definitively deliver a more suitable representation of the ground state wavefunction than any GRU variant could. On the other hand, we cannot be sure that the surprising underperformance of Lorentz and Poincar\'e GRU in these experiments compared to Lorentz RNN is an actual fact or  an artefact of their spatial constraint hyperparameter not being optimally selected.
\\\\
In terms of the averaged performance of each NQS ansatz obtained from taking the average of all VMC runs (see Fig.\ref{j1j2j3-all-comparison-ave}), the emerging picture in ascending order is as follows.
\ben
 \item $(J_2, J_3) = (0.0, 0.5)$:  Euclidean RNN (sixth),  Euclidean GRU  (fifth),  Poincar\'e RNN  (fourth), Lorentz GRU (third),  Poincar\'e GRU (second), Lorentz RNN  (first). 
\item $(J_2, J_3) = (0.2, 0.2)$: Euclidean RNN (sixth),  Euclidean GRU  (fifth),  Poincar\'e RNN  (fourth), Poincar\'e GRU  (third),  Lorentz GRU (second), Lorentz RNN  (first). 
\item $(J_2, J_3) = (0.2, 0.5)$: Euclidean RNN (sixth), Poincar\'e RNN  (fifth), Lorentz GRU (fourth),  Poincar\'e GRU (third), Euclidean GRU (second), Lorentz RNN  (first). 
\item $(J_2, J_3) = (0.5, 0.2)$: Euclidean RNN (sixth), Poincar\'e RNN  (fifth), Poincar\'e GRU (fourth),  Euclidean GRU (third),  Lorentz GRU (second), Lorentz RNN  (first). 
\enn
As in the case of $J_1J_2$ model above, the difficulties and instability in training hyperbolic NQS ans\"atze persisted in this case as well. In particular,  Lorentz GRU and RNN NQS failed to converge for all values of $J_2$ with some of the random seeds (see Tables \ref{full-j1j2j3-vmc-lrnn} and \ref{full-j1j2j3-vmc-lgru}). When they did converge, the values actually stayed in a narrow range without significant deviations.  As explained in detail in the previous $J_1J_2$ case, this behavior is expected since the hyperbolic optimization landscape is much harder to navigate compared to a flat, Euclidean one consisting only of flat directions.  Interestingly, despite this hyperbolic optimization challenge, both Poincar\'e GRU and RNN NQS (see Tables \ref{full-j1j2j3-vmc-prnn} and \ref{full-j1j2j3-vmc-pgru}) showed much more stability than the Lorentzian hyperbolic counterparts with only a single instance of non-convergence when $(J_2, J_3)=(0.5,0.2)$ in all $J_1J_2J_3$ experiments. Furthermore, Poincar\'e RNN/GRU also yielded very consistent mean energy values across mulitple random seeds.
\\\\
Conversely, for Euclidean GRU/RNN NQS ans\"atze (see Tables \ref{full-j1j2j3-vmc-ernn} and \ref{full-j1j2j3-vmc-egru}) which are supposedly more stable to train, we did run into the issues of inconsistent performances across different random seeds. One example of this occurred in the $J_2=0.0, J_3=0.5$ case with Euclidean GRU NQS, which yielded a mean energy in the range [-37, -38] for three of five seeds, with one seed failing to converge and the last seed performing particularly well and obtaining -53.29. Another example also involves Euclidean GRU failing to converge on one instance when $(J_2,J_3) = (0.5, 0.2)$. Stability-wise, we observe that while Euclidean GRU NQS failed to converge on two instances, Poincar\'e GRU and RNN NQS only failed to converge on one occasion, meaning that the Poincar\'e hyperbolic variants were actually more stable in training than the Euclidean GRU NQS in this set of $J_1J_2J_3$ experiments.
\\\\
Furthermore, when the averaged results are taken into account, it is interesting to see that Poincar\'e RNN did outperform Euclidean GRU on two occasions when $(J_2,J_3)$=(0.0, 0.5), (0.2, 0.2). Lorentz RNN, in particular, still ranked first despite its inherent instability in the training process across different seeds because when it did converge, the values are actually very consistently close to the true ground state energy with only minor fluctuations.
\end{itemize}

\begin{table}[!ht]
\centering
\begin{tabular}{lcc ccc}
\hline\hline
Ansatz &   (0.0, 0.5)  &  (0.2, 0.2) & (0.2,0.5) & (0.5,0.2) &
\\\hline\hline
Euclidean RNN  & -37.6130  & -25.3654  &  -32.6716   & -18.0043 &\\
                & 0.0022 &   0.0009 & 0.0052 &  0.0013 &\\ \hline
Poincare RNN & -50.9784 & -41.7109  &  -46.8415  &  -37.2335 &\\
                &0.0270 & 0.0177 &  0.0250 &  0.0090 &\\\hline
Lorentz RNN &\textbf{-53.9317}   &\textbf{-43.4314}    &\textbf{-49.5567}    &\textbf{-38.3244} &\\
                &0.0306 & 0.0058 & 0.0083 & 0.0042 &\\
\hline
Euclidean GRU &  -53.2965&   -42.8726  &  -48.9555   & -37.7999 &\\
                &0.0161 & 0.0059  & 0.0140 & 0.0278&\\\hline
Poincare GRU  &  -53.5432 &  -42.8895&  -48.9895   & -38.3142 &\\
                &0.0201 & 0.0083 & 0.0145 & 0.0052&\\\hline
Lorentz GRU & -53.8242  & -43.1327   & -49.1531  &  -38.0665 &\\
              & 0.0094 &  0.0084 & 0.0073 & 0.0044 &
\\\hline
DMRG& -53.9914 & -43.5860  &  -49.6287  &  -38.5473&\\
\hline
\end{tabular}{}
\caption{The best achievable VMC results, chosen from different VMC runs corresponding to different random seeds, of 1D $J_1J_2J_3$ Heisenberg models, where $J_1 = 1.0$ and $(J_2, J_3) = (0.0, 0.5)$,$(0.2, 0.2)$, $(0.2, 0.5)$,($0.5,0.2)$,  for all six NQS ans\"atze. For each ansatz, the mean energy is recorded in the first line, followed by the standard error directly below in the second line. All GRU variants have the same number of 15614 parameters. The number of samples used for inference is 10000. The best results are noted in bold.} \label{j1j2j3-RNN-GRU}
\end{table}
\FloatBarrier
\begin{table}[!ht]
\centering
\begin{tabular}{lcc ccc}
\hline\hline
Ansatz &   (0.0, 0.5)  &  (0.2, 0.2) & (0.2,0.5) & (0.5,0.2) &
\\\hline\hline
Euclidean RNN  & -37.1476 &  -25.2452  &  -32.2730  &  -17.6701&\\
                & 0.0020 &   0.0030 & 0.0007 & 0.0034 &\\ \hline
Poincare RNN & -50.7758  &-41.6542  &  -46.3494   & -36.9660 &\\
                &0.0275 & 0.0178 &   0.0267 &  0.0133 &\\\hline
Lorentz RNN &\textbf{-53.7652} &   \textbf{-43.4264}   & \textbf{ -49.4371}  &  \textbf{-38.3119} &\\
                &0.0146 & 0.0045 &  0.0109 &  0.0074 &\\
\hline
Euclidean GRU & -41.6693 &   -40.0682   & -48.6910 &   -37.4602 &\\
                &0.0276 & 0.0089  & 0.0163 &   0.0177 &\\\hline
Poincare GRU  &-53.1828 &  -42.5496 &   -48.3761 &   -37.2290 &\\
                & 0.0238   & 0.0100 & 0.0151 & 0.0152&\\\hline
Lorentz GRU & -52.5449  & -42.6764 &   -48.3182  &  -37.7768 &\\
                &0.0191 & 0.0065&   0.0150 &  0.0046 &
\\\hline
DMRG& -53.9914 & -43.5860  &  -49.6287  &  -38.5473&\\
\hline
\end{tabular}{}
\caption{The VMC results of 1D $J_1J_2J_3$ Heisenberg models - averaged from different runs with different random seeds, where $J_1 = 1.0$ and $(J_2, J_3) = (0.0, 0.5)$,$(0.2, 0.2)$, $(0.2, 0.5)$,($0.5,0.2)$,  with three variants of GRU ans\"atze: Euclidean GRU, Poincar\'e hyperbolic GRU and Lorentz hyperbolic GRU. For each variant, the mean energy is recorded in the first line, followed by the standard error directly below in the second line. All GRU variants have the same number of 15614 parameters. The number of samples used for inference is 10000. The best results are noted in bold.} \label{j1j2j3-RNN-GRU-ave}
\end{table}
\FloatBarrier
\FloatBarrier
\clearpage
\FloatBarrier
\clearpage

\section{Summary and future directions} \label{sum-concl}
In this work, we constructed novel types of hyperbolic NQS ans\"atze based on the additional types of hyperbolic RNNs including Poincar\'e RNN, Lorentz RNN, and Lorentz GRU, alongside the Poincar\'e GRU NQS which was first constructed in our previous work \cite{hld-hypnqs-25}. On the one hand, the formulations of the Lorentz RNN/GRU as standalone Lorentz recurrent architectures are the first constructions of their kinds and constitute one of the main novelties of this work.
On the other hand, in introducing these new types of hyperbolic NQS ans\"atze and extending the experiments of \cite{hld-hypnqs-25} to the $J_1J_2$ and $J_1J_2J_3$ Heisenberg models with 100 spins (twice the number of spins considered for the $J_1J_2$ model in \cite{hld-hypnqs-25} and three times the system size considered for $J_1J_2J_3$ in \cite{hld-hypnqs-25}), we not only reproduced and confirmed the results of \cite{hld-hypnqs-25} showing that Poincar\'e GRU always outperforming Euclidean GRU for all the $J_2$ and $(J_2, J_3)$ couplings studied, but also generalized these results to include the three new types of hyperbolic NQS introduced here.
\\\\
Given the different varieties of the six NQS ans\"atze studied in this work, which span two geometries (Euclidean versus hyperbolic) including two hyperbolic sub-geometries (Poincar\'e disk versus Lorentz hyperboloid) and two different neural network architectures (RNN versus GRU),  different kinds of performance comparisons can be made based on either the criterion of geometry, architecture or both. In what follows, we will use the results of our VMC experiments - collectively, across the eight different experiment settings corresponding to Heisenberg $J_1J_2$ model with $J_2=0.0$, 0.2, 0.5, 0.8 and Heisenberg $J_1J_2J_3$ model with ($J_2,J_3$)=(0.0, 0.5), (0.2,0.2), (0.2, 0.5), (0.5, 0.2) - to make the comparisons based on all possible combinations of criteria, in tems of both the best obtained and averaged results from multiple VMC runs using different random seeds. Since the `best achievable' result of each NQS ansatz (chosen from different VMC runs) represents its absolute capacity while the averaged results represent the stability of the ansatz, we will focus more on the former to make the comparisons, as we are ultimately interested in the best values reachable by each ansatz.  To facilitate the comparisons, we listed in Table \ref{tab:j1j2-top3}  and Table \ref{tab:j1j2j3-top3} the top three best performing NQS ans\"atze (in terms of best achievable mean energy) for each of the eight experiments\footnote{where one experiment corresponding to either a single value of $J_2$ or the pair $(J_2, J_3)$ consists of at least 30 different VMC runs where each of the 6 NQS ans\"atze was used with 5 different random seeds. This is not inclusive of the many more VMC runs required to tune the various hyperparameters of the hyperbolic Poincar\'e and Lorentz NQS in order to select the optimal spatial constraint parameters $R_{max}$ and $L_{max}$}.
\begin{table}[!h]
\centering
\begin{tabular}{c cccc c}
\hline\hline
 & $J_2=0.0$ & $J_2=0.2$ & $J_2=0.5$ & $J_2=0.8$\\
\hline\hline
 & $\begin{array}{c} \text{Lorentz RNN} \\ \text{Lorentz GRU}\\ \text{Poincar\'e GRU}\end{array}$
&  $\begin{array}{c} \text{Lorentz RNN} \\\text{Poincar\'e GRU} \\ \text{Lorentz GRU}\end{array}$
&  $\begin{array}{c}\text{Lorentz GRU}\\\text{Poincar\'e GRU}\\  \text{Euclidean GRU} \end{array}$
&  $\begin{array}{c}\text{Poincar\'e GRU}\\ \text{Lorentz GRU}\\ \text{Euclidean GRU} \end{array}$
\\
\hline\hline
\end{tabular}
\caption{The top three performing NQS ans\"atze in terms of best achievable mean energy, in descending order from top to bottom within each entry, for each $J_2$ coupling in the Heisenberg $J_1J_2$ VMC experiments. When $J_2=0.0$ and $J_2=0.5$, these top three lists include Lorentz RNN, Lorentz GRU and Poincar\'e GRU in different orders, while for $J_2=0.5$ and $J_2=0.8$ the top three spots consisted of the GRU variants including Lorentz, Poincar\'e and Euclidean GRU. } \label{tab:j1j2-top3}
\end{table}

\begin{table}[!h]
\centering
\begin{tabular}{c cccc c}
\hline\hline
 & $(0.0,0.5)$ & $(0.2,0.2)$ & $(0.2,0.5)$ & $(0.5,0.2)$\\
\hline\hline
 & $\begin{array}{c} \text{Lorentz RNN} \\ \text{Lorentz GRU}\\ \text{Poincar\'e GRU}\end{array}$
&  $\begin{array}{c} \text{Lorentz RNN} \\ \text{Lorentz GRU}\\\text{Poincar\'e GRU} \end{array}$
&  $\begin{array}{c}\text{Lorentz RNN}\\  \text{Lorentz GRU}\\\text{Poincar\'e GRU} \end{array}$
&  $\begin{array}{c}\text{Lorentz RNN}\\\text{Poincar\'e GRU}\\  \text{Lorentz GRU} \end{array}$
\\
\hline\hline
\end{tabular}
\caption{The top three performing NQS ans\"atze in terms of best achievable mean energy, in descending order from top to bottom within each entry, for each $(J_2,J_3)$ coupling pair in the Heisenberg $J_1J_2J_3$ VMC experiments. These four top three lists are made up entirely of Lorentz RNN, Lorentz GRU and Poincar\'e GRU, in different orders. } \label{tab:j1j2j3-top3}
\end{table}
\begin{enumerate}
\item Between hyperbolic and Euclidean NQS ans\"atze of the \textit{same} architecture (in other words, comparing Euclidean versus hyperbolic geometry):
\begin{enumerate}  
    \item Poincar\'e RNN and Lorentz RNN, the newly introduced hyperbolic RNN NQS ans\"atze in this work, always outperformed Euclidean RNN in all experiments (including $J_2=0.0$).
    \item Lorentz GRU - the newly introduced hyperbolic GRU NQS ansatz in this work, and Poincar\'e GRU - the previously introduced hyperbolic GRU NQS ansatz from \cite{hld-hypnqs-25}, always outperformed Euclidean GRU, in all experiments (including $J_2=0.0$).
\end{enumerate}
These observations point towards the fact that within the same neural network architecture (where the NQS ans\"atze have the same number of parameters), the Poincar\'e and Lorentz hyperbolic variants (either RNN or GRU) might be potentially  superior to and very \textit{likely} outperform the corresponding Euclidean variant (either RNN or GRU) in this problem of approximating the ground state wavefunction of Heisenberg spin systems with varying hierarchical degrees of nearest neighbor couplings.
With these, the primary goal of this work, which concerns the generalizations of our previous results reported in \cite{hld-hypnqs-25}, which only included two NQS ans\"atze - the Euclidean and Poincar\'e GRU -  in the $J_1J_2$ and $J_1J_2J_3$ models, to all four hyperbolic NQS variants, is achieved.
\item  Between hyperbolic NQS ans\"atze of the same architecture (in other words, comparing the hyperbolic sub-geometry of Poincar\'e model versus Lorentz model):
\begin{enumerate}
    \item \textit{RNN architecture}: The Lorentz hyperboloid geometry wins here, with Lorentz RNN definitively outperformed Poincar\'e RNN, for all couplings considered in all experiments.
    \item \textit{GRU architecture}:  Poincar\'e GRU outperformed Lorentz GRU three out of eight times - twice  in $J_1J_2$ experiments ($J_2=0.2,0.8)$, and once in $J_1J_2J_3$ experiments when $(J_2,J_3) = (0.5,0.2)$. Conversely, Lorentz GRU  outperformed Poincar\'e GRU five out of eight times, twice in $J_1J_2$ experiments when $J_2=0.0, 0.5$, and thrice in $J_1J_2J_3$ experiments when $(J_2,J_3) = (0.0, 0.5)$, (0.2,0.2), (0.2,0.5).
\end{enumerate}
These observations point towards the likelihood that the Lorentz model of hyperbolic space is potentially superior to the Poincar\'e model, given the number of times the Lorentz variant collectively outperformed the Poincar\'e variant across two different architectures. Furthermore, our observations here also agree with those from the literature involving studies done on NLP \cite{hyprnn-lorentz-1705}, \cite{hyprnn-lorentz-1806} which found that the Lorentz model of hyperbolic space tended to be better than the Poincar\'e disk model due to its open nature that is not restricted to the hard edge at $R=1$.
However, in the quantum many-body systems, the relative performances of the Lorentz NQS variants versus the Poincar\'e NQS variants require further studies as we could not be certain that the optimal choices of $R_{max}$ for Poincar\'e RNN/GRU or $L_{max}$ for Lorentz RNN/GRU were obtained in this work, as we relied on the trial-and-error process to pick the best performing value of $R_{max}$ and $L_{max}$ in each experiment, due to our limited computing resources which prevented a comprehensive scanning of parameters that could reliably pick the optimal values.
\item Between hyperbolic NQS ans\"atze of different architectures (in other words, comparing RNN versus GRU given the same hyperbolic sub-geometry):
\begin{enumerate}
\item \textit{Poincar\'e disk geometry}: Conforming to expectations, Poincar\'e RNN (5394/6964 parameters) always underperformed Poincar\'e GRU (15614 parameters) in all experiments,  due to the more sophisticated GRU gating mechanism of Poincar\'e GRU and the fact that Poincar\'e GRU had 2.25 to 3 times more parameters than Poincar\'e RNN. However, the performance gap between Poincar\'e RNN and Poincar\'e GRU is much smaller than the performance gap between Euclidean RNN and Euclidean GRU (where the former trailed the latter by a very large margin).
\item \textit{Lorentz hyperboloid geometry}: Surprisingly, while Lorentz GRU (15614 parameters) is expected to be the better hyperbolic variant due to its more sophisticated GRU gating mechanism, it only outperformed Lorentz RNN (5394/6964 parameters) a total two out of eight times, twice in $J_1J_2$ model when $J_2=0.5, 0.8$. 
On the remaining six occasions, when $J_2 = 0.0, 0.2$ in Heisenberg $J_1J_2$ model and all 4 values of $(J_2, J_3)$ couplings in $J_1J_2J_3$ model, Lorentz RNN outperformed Lorentz GRU. In all theses instances, despite the large gap in the number of parameters, the performance gap between Lorentz GRU and Lorentz RNN is very small (smaller than that between Poincar\'e RNN and Poincar\'e GRU), with Lorentz RNN always emerging as the better NQS. 
\end{enumerate}
Given this small performance gaps between Poincar\'e/Lorentz RNN and Poincar\'e/Lorentz  GRU despite the large gap in the number of parameters, it would be interesting to carry out more experiments involving the hyperbolic RNN/GRU with the same number of parameters to study their performances.
\item  \textit{Across different geometries and architectures}: This is the most general comparison  that aims to determine which model tends to have the best performance the most number of times. From Table \ref{tab:j1j2-top3} and Table \ref{tab:j1j2j3-top3} that list the top three performing NQS ans\"atze in each of the eight Heisenberg experiments, and the observations listed above, it is obvious that \textit{geometry-wise}, hyperbolic (either Lorentz or Poincar\'e) NQS ans\"atze always outperformed their Euclidean counterparts. As already mentioned above, this result actually fulfilled the primary goal of this work, which sets out to construct new hyperbolic NQS variants and generalize the main findings of \cite{hld-hypnqs-25} regarding the role of the underlying geometry in delivering the outperformance of Poincar\'e versus Euclidean GRU to all three new hyperbolic variants.
\\\\
With the main goal settled, we are left with a more nuanced picture when zooming in closer to inspect the performances of the four hyperbolic NQS variants relative to one another, primarily because of the lack of a systematic method - besides the computationally expedient trial-and-error method used - to select the optimal $R_{max}$ and $L_{max}$ hyperparameters. With our current selection of $R_{max}$ and $L_{max}$ as documented in Section \ref{sec-rmax} and Section \ref{sec-lmax}, the results we have obtained is listed below.
\begin{enumerate}
    \item Lorentz RNN ranked first six out of eight times, twice in the $J_1J_2$ experiments when  $J_2=0.0, 0.2$, and all four times in the $J_1J_2J_3$ experiments for all values of ($J_2,J_3$) couplings.
    \item Poincar\'e GRU ranked first one out of eight times, in the $J_1J_2$ experiments when $J_2=0.8$. 
    \item Lorentz GRU ranked first one out of eight times,  in the  $J_1J_2$ experiments when $J_2=0.5$.
\end{enumerate}
Since Lorentz RNN was the top performing variant the most number of times, it is the overall best hyperbolic variant in the studies performed in this work. However, as mentioned previously in the main text,  in those cases where $J_2$ or ($J_2, J_3$) couplings were small (0.0 or 0.2), it is possible that the pure hyperbolic geometry underlying Lorentz RNN could probably deliver a more suitable representation of the ground state wavefunction than either Poincar\'e GRU or Lorentz GRU could, despite their combination of both hyperbolic geometry and complex GRU gating mechanism. On the other hand, with larger couplings when $(J_2,J_3)$ = (0.5, 0.2) or (0.2, 0.5), it is very likely that Lorentz RNN outperforming either Poincar\'e GRU or Lorentz GRU has more to do with the latter two being suboptimally tuned (because of their $R_{max}$ and $L_{max}$ hyperparameters). To find out which of these possibilities is more likely, i.e. whether it is within the capability of Lorentz RNN to do better than both Lorentz/Poincar\'e GRU, or whether it is the poor selection of $R_{max}/L_{max}$ that rendered the latter two to underpderform, it is necessary to carry out these experiments with more comprehensive selections of the spatial constraint hyperparameters for all hyperbolic networks - a task we hope to return to in the future should we have more computational resources at our disposal.
However, regardless of whether Lorentz RNN could actually outperform Lorentz/Poincar\'e GRU, a noteworthy  result that came out of these experiments is that Lorentz RNN - with up to three times fewer parameters - did outperform Euclidean GRU six out of eight times. This makes it highly useful for future NQS studies of other quantum many-body systems.
\\\\
On the other hand, the main limitations in dealing with hyperbolic NQS in general and Lorentz RNN/GRU in particular are the long training times and training instabilities as seen from the VMC experiments done across multiple random seeds. Because of the more complicated mathematical operations in hyperbolic space that were used to construct the hyperbolic NQS ans\"atze, longer training times compared to Euclidean NQS cannot be prevented but can be mitigated with hardware accelerators. In terms of training stability,
we did observe that for the experiments conducted in this work, while Poincar\'e GRU (or even Poincar\'e RNN in the $J_1J_2J_3$ case) could be as good as Euclidean GRU when different random seeds were used, Lorentz GRU and RNN NQS displayed more unstable behaviors involving non-convergence issues on a few occasions for some of the $J_2$ and ($J_2,J_3$) couplings. These instabilities, well known from other works dealing with hyperbolic neural networks in the context of NLP, are directly attributable to the rugged hyperbolic optimization landscape that makes the hyperbolic Lorentz and Poincar\'e NQS much more challenging to train than Euclidean ones. Another relevant point of note is the fact that our Lorentz/Poincar\'e constructions have not made used of any nonlinear activation functions, instead we have relied solely on the inherent nonlinearity of the underlying hyperbolic geometry in a similar manner to the original construction of \cite{ganea-1805} in NLP works. Introducing an explicit, separate nonlinear activation function such as `\texttt{tanh}' or  `\texttt{elu}' into the constructions Eq.\ref{eq-hrnn}, Eq.\ref{eq-hgru}, Eq.\ref{eq-lorentz-rnn}, Eq.\ref{eq-lorentz-gru} could potentially regularize the output and make the training behaviors of Poincar\'e RNN and Lorentz RNN/GRU more stable, albeit at the cost of increased training time\footnote{as noted in the recent work \cite{hld-2dhyprnn} that the use of the nonlinear activation `\texttt{elu}' in the  two-dimensional hyperbolic Lorentz RNN construction led to an improvement in the performance of the Lorentz 2DRNN compared to the version without any nonlinear activation.}.
\end{enumerate}
In conclusions, given the results from this work concerning the newly constructed hyperbolic NQS ans\"atze including the Poincar\'e RNN/GRU and Lorentz RNN/GRU that validate and generalize those of \cite{hld-hypnqs-25}, namely the fact that hyperbolic NQS are likely to outperform their Euclidean counterparts in quantum many-body systems with structural hierarchy in the form of different degrees of neighbor interactions, we briefly sketch the main possible extensions and future directions below.
\bei
\item
The most direct extensions of this work would be the constructions of other types of autoregressive hyperbolic NQS ans\"atze beyond the recurrent neural network architecture and the verification of their performances in the paradigmatic Heisenberg systems. Of particular interest will be those hyperbolic NQS ans\"atze based on convolutional neural network or masked transformers, or even hybrid RNN-CNN  architectures (\cite{hld-cicy-rnn}), among other choices.
\item Another direct extension of this work and the work \cite{hld-hypnqs-25} would be to apply these newly constructed types of hyperbolic RNN NQS ans\"atze  to the problem of $N\times N$ two-dimensional Transverse Field Ising Model (TFIM) in a similar manner to what was done in \cite{hld-hypnqs-25}, where we showed that one-dimensional (1D) Poincar\'e GRU NQS definitively outperformed 1D Euclidean GRU NQS, because of the hierarchical structure of interactions between the first and $N^{th}$ nearest neighbors present when the two dimensional geometry of the problem is translated to one dimension. This was done very recently in the work \cite{hld-2dhyprnn} which not only studied the four variants of 1D hyperbolic NQS proposed in this work but also introduced the first type of two-dimensional hyperbolic NQS in the form of the Lorentz 2DRNN. A new and important insight originating from the work \cite{hld-2dhyprnn} is the observation that 1D and 2D hyperbolic NQS not only outperform Euclidean ones in the 2D TFIM quantum system exhibiting hierarchical structures (when translated to a one dimension setup) but also in those at criticality or phase transitions when the physics can be described by a conformal field theory (CFT) that is dual to an Anti-de-Sitter (AdS) space whose spatial geometry is hyperbolic, which matches with the hyperbolic geometry underlying the constructions of these hyperbolic NQS ans\"atze.
\item Another concrete extension of this work would be the conversion of the hyperbolic NQS constructions proposed in here and \cite{hld-hypnqs-25} to those with trainable curvatures, which would eliminate the need for using the spatial constraint hyperparameters $R_{max}$ and $L_{max}$, which have been added in a heuristic manner to avoid numerical instabilities in the training. A hyperbolic NQS with a trainable curvature would arguably be a more elegant and sturdier construction than the ones we have considered so far.
\item Further afield, more general directions regarding hyperbolic NQS exist, an interesting example of which is the variational studies of quantum systems with volume law entanglement property such as the maximally chaotic SYK (Sachdev-Ye-Kitaev) \cite{syk-1}, \cite{syk-2} model that is known to be the 1D toy model for a black hole. Theoretically at least, hyperbolic NQS might be a suitable tool to approximate the volume-law ground state of such a system, thanks to its exponentially growing volume capacity that can potentially handle the random all-to-all couplings of the SYK model in a way that other NQS or tensor methods might not have been able to do.
Another example of the many possible future directions is the deployment of these exact hyperbolic NQS ans\"atze in quantum systems beyond those of Heisenberg and TFIM, perhaps those that arise from a high energy physics context like the bosonic and supersymmetric matrix models that have been studied using variational quantum circuits \cite{hld-qc}. It might be instructive, in particular, to study  the supersymmetric matrix models with hyperbolic NQS in a VMC setting because, very broadly speaking, some of these systems arise from the dimensional reduction of super Yang-Mills theories which can be dual to AdS spaces (via the AdS/CFT correspondance \cite{ads-cft}). As pointed out in \cite{hld-2dhyprnn}, hyperbolic NQS ans\"atze are particularly well-suited to approximate the ground states of those systems describable by CFT physics because of the match in the hyperbolic geometry underlying the constructions of these NQS and the spatial section of AdS space.
\eni
We hope to return to some of the listed directions above in future works.

\clearpage
\section{Appendix}
\subsection{Spatial constraint hyperparameters for hyperbolic NQS} \label{spatial-constr}
In the VMC experiments done in this work, both Poincar\'e RNN/GRU and Lorentz RNN/GRU required some form of spatial constraint in order to attain their optimal performances. For Poincar\'e networks, this is the maximal radius in the Poincar\'e disk, denoted as $R_{max}$. For Lorentz networks, this is the maximal spatial norm that the spatial components of the Lorentz vector are not allowed to exceed - recall that in the Lorentz hyperboloid model, the time component $x_0$ and spatial components $\vec x$ of the vector $\vec{\textbf x}= (x_0, \vec x)$ are constrained by the defining equation $-x_0^2 + \vec x\,.\,\vec x = -1$, so it suffices to impose a spatial norm on the spatial components of the Lorentz vector $x$, which automatically constrains the time component as well. We note that imposing these spatial contraint hyperparameters on the Poincar\'e and Lorentz hyperbolic NQS is equivalent to imposing an uppber bound on the curvature of the underlying hyperbolic space (either the Poincar\'e disk or the Lorentz hyperboloid). While this has proved to be useful in these experiments conducted in this work, a more elegant approach might involve the conversion of the fixed curvature $k$ of the underlying hyperbolic space into a trainable parameter for the hyperbolic NQS.
\subsubsection{Effects of maximal radius on Poincar\'e hyperbolic RNN/GRU} \label{sec-rmax}
\begin{figure}[!h]
\centering
\includegraphics[width = .7\textwidth]{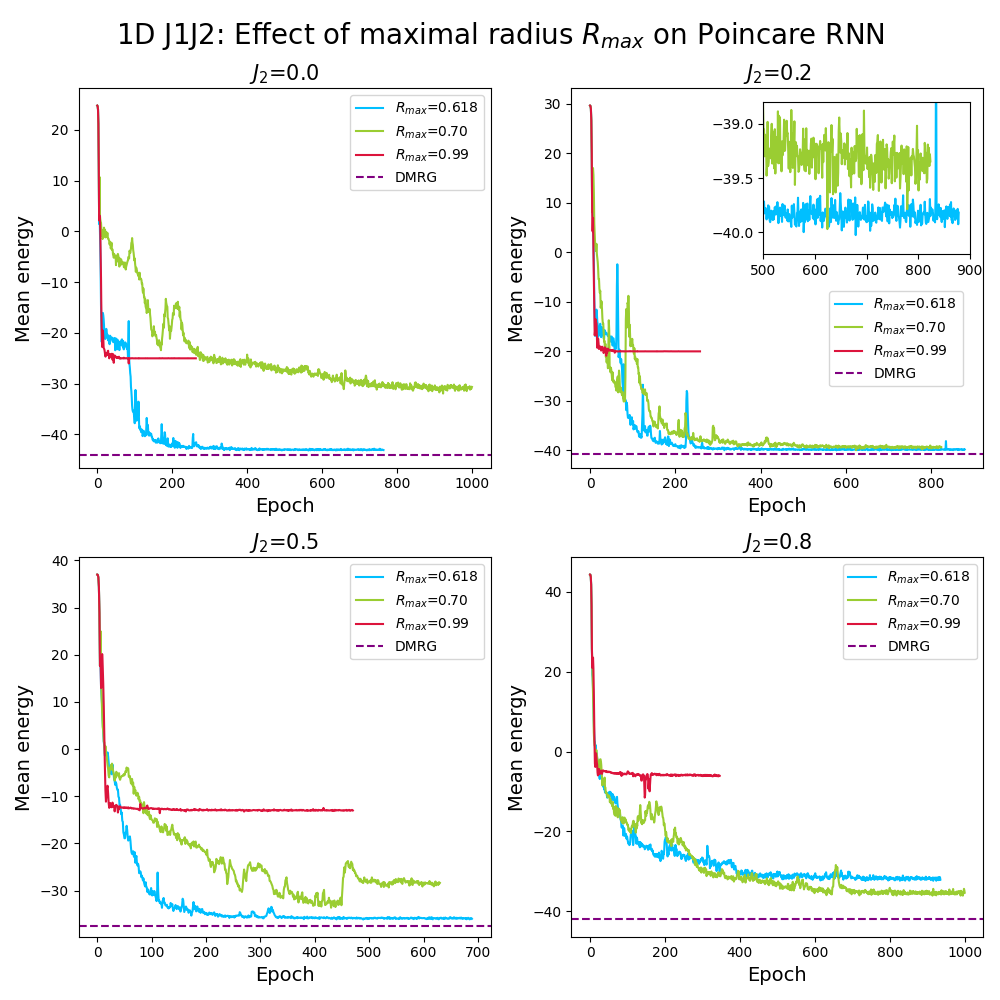}
\caption{An illustration of the effect of the maximal radius $R_{max}$ in the Poincar\'e disk on the performance of Poincar\'e RNN for the  1D Heisenberg $J_1J_2$ model with $J_1 = 1.0$ and different $J_2$ couplings - clockwise $J_2 = 0.0, 0.2, 0.8, 0.5$ with the fixed random seed of 111. An optimal $R_{max}$ must be chosen to ensure the optimal performance of the Poincar\'e RNN. When $R_{max}=0.99$ (nearing the edge of the Poincar\'e disk), due to numerical explosion, the performance of Poincar\'e RNN is always the worst - comparable to that of the Euclidean RNN. Due to the automatic early stopping mechanism upon no registered improvements, the number of epochs differed for each training run.  When present, the inset plot within each plot shows a magnified segment towards the end of the training when the models neared convergence. These inset plots are only included when the curves are not sufficiently distinct in the main plots. }\label{j1j2-r_max_prnn-curves}
\end{figure}

The effect of $R_{max}$ is most pronounced on the performances of Poincar\'e RNN, as illustrated in Fig.\ref{j1j2-r_max_prnn-curves} in the $J_1J_2$ case, although the situation is similar in the $J_1J_2J_3$ case. Without any $R_{max}$ (or equivalently $R_{max}$ is by default 1.0, at the edge of the Poincar\'e disk), Lorentz RNN performance was almost identical to that of Euclidean RNN, which converged to a local minimum much higher than the true ground state. During these trainings with no $R_{max}$, the norms of the hidden state $\vec h_i$ consistently stayed high at 0.9999 or higher, meaning that a numerical explosion had already occurred, causing the model to lose the ability to learn. When lower $R_{max}$ values were used, the situation improved dramatically (see Fig.\ref{j1j2-r_max_prnn-curves}), resulting in the significance improvement seen in the performances of Poincar\'e RNN.
\\\\
On the other hand, Poincar\'e GRU is much more stable to train than Poincar\'e RNN, thanks to its complex gating mechanism that updates the final hidden state $\vec h_i$ using a combination of old $(\vec h_{i-1})$ and new information ($\vec{\tilde h}_i$), determined by the value of the update gate $\vec z_i$. As such, the use of $R_{max}$ was not necessary for Poincar\'e GRU in the 1D $J_1J_2$ model when the default value of $R_{max}=1.0$ (the edge of the Poincar\'e) disk was used for the first three couplings $J_2=0.0, 0.2, 0.5$. When $J_2=0.8$, however, we found that restricting $R_{max}<1$ led to an improvement in the performance of Poincar\'e GRU NQS. For the $J_1J_2J_3$ model, $R_{max}<1$ was necessary to obtain stable training behaviors for $(J_2,J_3)$ = (0.2, 0.5) and (0.5, 0.2). We also note that with the same $R_{max}$ parameter, different random seeds yielded different training outcomes. In most instances\footnote{where an instance here refers to the combination of $J_2$ and the NQS ansatz under consideration}, the results were broadly similar across different seeds but in some cases, the same $R_{max}$ could lead to significantly different results with different seeds. This is actually statistically expected, especially when dealing with hyperbolic NQS with the much more rugged optimization landscape, since each random seed represents a different initialization configuration which was known to result in vastly different optimization/training outcome even for Euclidean RNN/GRU NQS with much easier optimization landscape\footnote{It should be noted that with different hardwares, the same seed might not produce the same result but a collection of the same seeds should produce the same statistical behavior. This is best explained using a concrete example: Consider Euclidean RNN NQS when $J_2=0.5$, which almost always converges to the false minimum of around $\approx$-13 for about 80-90 percent of the time, with a 10-20 percent chance of converging at a much lower value of $\approx$-35. On two different machines, using 5 seeds (111,222,333,444,555), seed 333 might be the one that yields $\approx$-35 on one machine while seed 555 might be the one to yield $\approx$-35 on another machine, while the rest of the seeds consistently obtain the value that is within a narrow range of  $\approx$-13, but at least one of the 5 seeds considered would yield $\approx$-35.}.
\\\\
The choices of $R_{max}$ used for Poincar\'e RNN \& GRU for different $J_2$ couplings in Heisenberg $J_1J_2$ model and for ($J_2,J_3$) couplings in Heisenberg $J_1J_2J_3$ model are recorded in Table \ref{j1j2-hyperparams-poincare} and Table \ref{j1j2j3-hyperparams-poincare}, respectively. Due to a lack of computational resources, these values were arrived at by a process of trial-and-error, in which we only tested a small number of $R_{max}$ values, rather than by using a comprehensive scanning algorithm, which means that it is highly likely that these values are not the absolutely optimal values that could guarantee the best performances possible by the Poincar\'e RNN and GRU.
\begin{table}[!ht]
\centering
\begin{tabular}{lcc ccc}
\hline\hline
Ansatz &  $J_2 = 0.0$  & $J_2 = 0.2$ & $J_2 = 0.5$ & $J_2 = 0.8$ &
\\\hline\hline
Poincar\'e RNN  &   0.618  & 0.618   &  0.8&  0.95&\\
Poincar\'e GRU & 1.0 & 1.0 & 1.0& 0.8
\\\hline
\hline
\end{tabular}{}
\caption{The maximal radius $R_{max}$ in the Poincar\'e disk chosen for Poincar\'e GRU and RNN NQS ans\"atze for the $J_1J_2$Heisenberg models, where $J_1 = 1.0$ and $J_2 = 0.0, 0.2, 0.5, 0.8$.  For Poincar\'e RNN, we tested $R_{max}=(0.618, 0.7, 0.99)$ for $J_2=0.0, 0.2$. For $J_2=0.5$, we tested $R_{max}=(0.618, 0.7, 0.8, 0.99)$. For $J_2=0.8$, we tested $R_{max} = (0.618, 0.7, 0.95, 0.99)$. Note that our choice of $R_{max}=0.618$ in the Poincar\'e model was motivated by the spatial norm $L_{max}= 2.0$ chosen for Lorentz RNN. For Poincar\'e GRU, we tested $R_{max}=(0.8, 1.0)$ when $J_2=0.8$.} \label{j1j2-hyperparams-poincare}
\end{table}
\begin{table}[!ht]
\centering
\begin{tabular}{lcc ccc}
\hline\hline
Ansatz &  $(0.0, 0.5)$  & $(0.2, 0.2)$ & $(0.2,0.5)$ & $(0.5,0.2)$ &
\\\hline\hline
Poincar\'e RNN  &  0.78 &  0.78  &  0.78 &  0.78 &\\
Poincar\'e GRU & 1.0 & 1.0 & 0.95& 0.7
\\\hline
\hline
\end{tabular}
\caption{The maximal radius $R_{max}$ in the Poincar\'e disk chosen for Poincar\'e GRU and RNN NQS ans\"atze for the  $J_1J_2J_3$ Heisenberg models with  $(J_2, J_3) = (0.0,0.5)$, $(0.2,0.2)$, $(0.2, 0.5)$, $(0.5,0.2)$. For Poincar\'e RNN, we tested $R_{max}=(0.99, 0.78)$ and then fixing $R_{max}0.78$. For Poincar\'e GRU, we tested $R_{max}=(0.7, 0.78, 0.82, 0.95)$. } \label{j1j2j3-hyperparams-poincare}
\end{table}
\FloatBarrier
\clearpage
\newpage
\subsubsection{Effects of maximal spatial norm on Lorenz hyperbolic RNN/GRU} \label{sec-lmax}
\begin{figure}[!h]
\centering
\includegraphics[width = .7\textwidth]{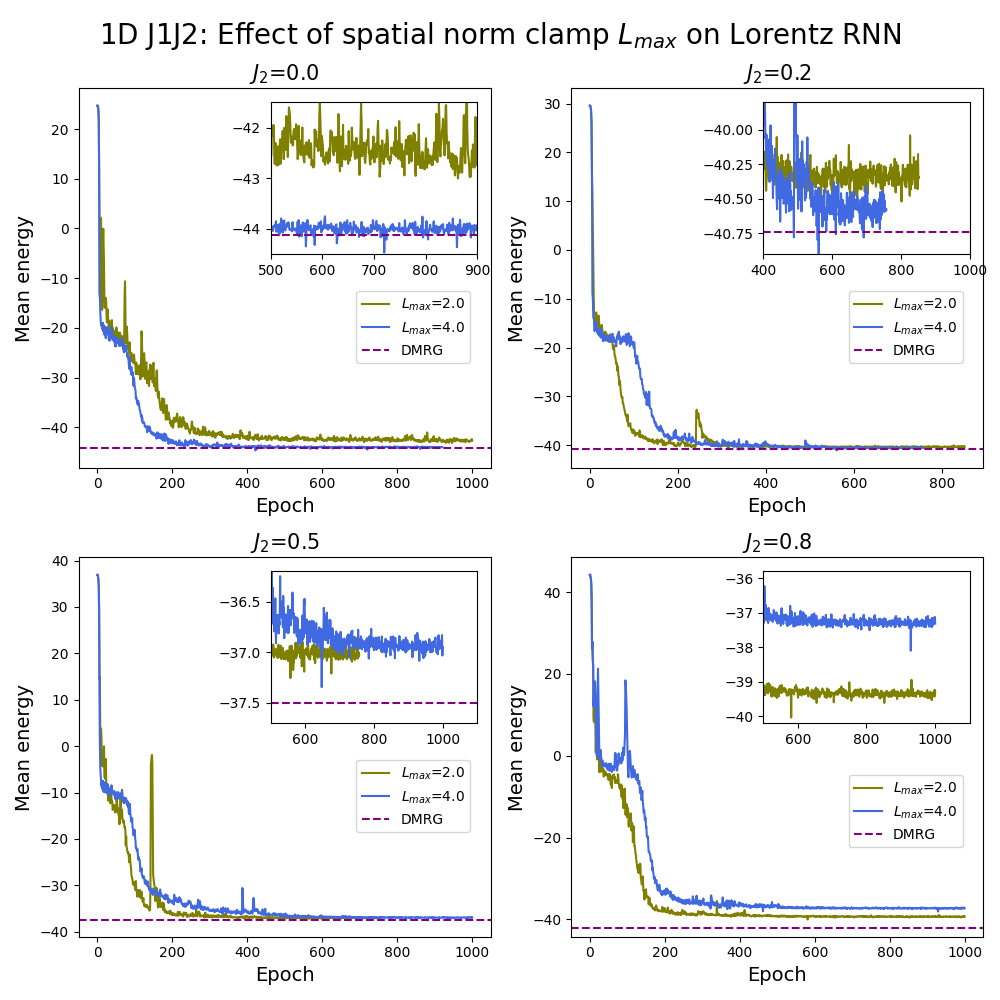}
\caption{An illustration of the effect of the spatial norm clamp $L_{max}$ in the Lorentz hyperboloid on the performance of Lorentz RNN for the  1D Heisenberg $J_1J_2$ model with $J_1 = 1.0$ and different $J_2$ couplings - clockwise $J_2 = 0.0, 0.2, 0.8, 0.5$ for the fixed random seed of 111. An optimal $L_{max}$ must be chosen to ensure the optimal performance of the Lorentz RNN.  Due to the automatic early stopping mechanism upon no registered improvements, the number of epochs differed for each training run.  When present, the inset plot within each plot shows a magnified segment towards the end of the training when the models neared convergence. These inset plots are only included when the curves are not sufficiently distinct in the main plots. }\label{j1j2-sc_lrnn-curves}
\end{figure}

The Lorentz hyperboloid model of hyperbolic space is an open model with no restrictions on the norm of the Lorentz vectors. As such, both Lorentz RNN and GRU are more prone to numerical instabilities and explosions than Poincar\'e RNN/GRU.  When this happens, the spatial norm of the components of the final hidden state $\vec h_i$ of the Lorentz RNN/GRU can reach very large values (exceeding 100 or 1000) during the training process.
It was absolutely necessary, at least for the experiments done in this work, to impose a maximal norm value $L_{max}$ on the spatial components (which also constrained the time component thanks to defining equation of the Lorentz hyperboloid) of the hidden state vector. Depending on the coupling regimes of interest (different values of $J_2$ or ($J_2,J_3$)), a smaller $L_{max}$ might be more suitable than a larger value of $L_{max}$, as illustrated in Fig. \ref{j1j2-sc_lrnn-curves} which shows the effect of different $L_{max}$ on the performances of Lorentz RNN. As noted previously, smaller $L_{max}$ values roughly correspond to flatter curvatures while higher $L_{max}$ values roughly correspond to larger curvatures.
\\\\
The choices of $L_{max}$ used for Lorentz RNN \& GRU for different $J_2$ couplings in Heisenberg $J_1J_2$ model and for ($J_2,J_3$) couplings in Heisenberg $J_1J_2J_3$ model were recorded in Table \ref{j1j2-hyperparams-lorentz} and Table \ref{j1j2j3-hyperparams-lorentz}, respectively. Similar to the Poincar\'e case above, due to a lack of computational resources, these values were arrived at by a process of trial-and-error, in which we tested a very small number of $L_{max}$ values, rather than by using a comprehensive scanning algorithm, which means that it is highly likely that these values are not the best possible values that could guarantee the best performances possible by Lorentz RNN and GRU.
\begin{table}[!ht]
\centering
\begin{tabular}{lcc ccc}
\hline\hline
Ansatz &  $J_2 = 0.0$  & $J_2 = 0.2$ & $J_2 = 0.5$ & $J_2 = 0.8$ &
\\\hline\hline
Lorentz RNN  &  4.0 (double) &  2.0 (double)  &  2.0 (double) &  2.0 (double) &\\
Lorentz GRU & 4.0 (single) & 6.0 (single) & 4.0 (single) & 4.0 (single)
\\\hline
\hline
\end{tabular}{}
\caption{The maximal spatial norm $L_{max}$ used in the Lorentz hyperboloid chosen for Lorentz GRU and RNN NQS ans\"atze for the  $J_1J_2$ Heisenberg models, where $J_1 = 1.0$ and $J_2 = 0.0, 0.2, 0.5, 0.8$. In brackets, `single' and 'double' refers to the number of times the spatial norm clamp is applied within the \texttt{forward} pass where the mathematical operations defining the LorentzRNN/GRU are coded, with `single' clamp being applied only on the hidden state at the beginning of the \texttt{forward} pass, and `double' clamps being applied to the state both at the beginning and the end of the pass.  For Lorentz RNN, we tested $L_{max}=(2.0, 4.0)$, for Lorentz GRU, we tested $L_{max}=(2.0,4.0,6.0)$, with different combinations of single and double clamps.} \label{j1j2-hyperparams-lorentz}
\end{table}
\begin{table}[!ht]
\centering
\begin{tabular}{lcc ccc}
\hline\hline
Ansatz &  $(0.0, 0.5)$  & $(0.2, 0.2)$ & $(0.2,0.5)$ & $(0.5,0.2)$ &
\\\hline\hline
Lorentz RNN  &  4.0 (double) &  4.0 (double)  &  4.0 (double) &  4.0 (double) &\\
Lorentz GRU & 4.0 (single) & 4.0 (single) & 4.0 (single) & 4.0 (single)
\\\hline
\hline
\end{tabular}
\caption{The maximal spatial norm $L_{max}$ used in the Lorentz hyperboloid chosen for Lorentz GRU and RNN NQS ans\"atze for the $J_1J_2J_3$ Heisenberg models where $J_1 = 1.0$ and $(J_2, J_3) = (0.0,0.5)$, $(0.2,0.2)$, $(0.2, 0.5)$, $(0.5,0.2)$. In brackets, `single' and 'double' refers to the number of times the spatial norm clamp is applied within the \texttt{forward} pass where the mathematical operations defining the LorentzRNN/GRU are coded, with `single' clamp being applied only on the hidden state at the beginning of the \texttt{forward} pass, and `double' clamps being applied to the state both at the beginning and the end of the pass.  For both Lorentz RNN/GRU, we fixed $L_{max}=4.0$ and tested that with different combinations of single and double clamps.} \label{j1j2j3-hyperparams-lorentz}
\end{table}\label{j1j2j3-hyperparams-lorentz}

\FloatBarrier
\clearpage
\subsection{Full results of VMC experiments with different random seeds} 
\label{res-random-seeds}
\subsubsection{$J_1J_2$ experiments} \label{full-res-j1j2}
In this section, we list the full results of all individual Heisenberg $J_1J_2$ VMC experiment run where one run corresponds to a single value of the $J_2$ coupling with a single NQS ansatz at a fixed random seed. For each NQS ansatz at a fixed $J_2$ value, we performed at least 5 runs with 5 different random seeds (111, 222, 333, 444, 555) while keeping all other hyperparameters such as Euclidean (and hyperbolic) learning rates, spatial constraint hyperparameters fixed. To select suitable hyperparameters for each ansatz, the 5-seed experiment runs have to be repeated with different ranges of the hyperparameters to tune. The results below are those obtained with the optimally selected hyperparameters. In some of the runs, for all NQS ans\"atze including both Euclidean and hyperbolic variants, we obtained non-converged results where the mean energy produced by the  NQS ansatz drifted around without settling into a recognizable minimum or the convergence was at a false minimum that is much higher (e.g. positive energy) than the actual true ground state energy. These non-converged results are not taken into account when calculating the averaged values reported in Table \ref{j1j2-RNN-GRU-ave}.
\begin{table}[!h]
\centering
\begin{tabular}{c ccccc c}
\hline\hline
 &  Seed  & $J_2=0.0$ &   $J_2=0.2$  &   $J_2=0.5$ &   $J_2=0.8$ &
\\ \hline\hline               
& 111 &-24.9966  &-19.9791 &   -13.0651  &  -16.5575 &\\
&& 0.0008 &  0.0059 & 0.0033 & 0.0389  &\\
\hline
& 222 & -25.1815  &  \textbf{-20.4812}   & -13.2436   & -5.9942  &\\
&& 0.0018  & 0.0018 & 0.0022&  0.0046 & \\ \hline
& 333 & -24.9222  &  -20.2627   & \textbf{-35.0857}   & \textbf{-24.5784} &\\
&& 0.0080  & 0.0052  & 0.0229 & 0.0436  &\\ \hline
& 444 &    \textbf{-25.2492}  &  -20.4136   & -12.2276  &  -20.1815 &\\
&& 0.0047   & 0.0029 &  0.0043 &  0.0024 &\\ \hline
& 555 &-24.9501 & -19.8035  &  -13.2426 &    -20.1285 &\\
&& 0.0031 &  0.0070 & 0.0020 & 0.0051& \\ \hline
& DMRG &  -44.1277 & -40.7388 &   -37.5000 &   -42.0701 &
\\ \hline
\end{tabular}
\caption{The full results of the Heisenberg $J_1J_2$ VMC experiments with $J_1=1.0$ and different $J_2$ couplings, corresponding to different random seeds for  Euclidean RNN NQS. For each entry the energy is listed first in the first line, followed by the standard error in the second row. The best results are noted in bold.}\label{full-j1j2-vmc-ernn}
\end{table}
\begin{table}[!h]
\centering
\begin{tabular}{c ccccc c}
\hline\hline
 &  Seed  & $J_2=0.0$ &   $J_2=0.2$  &   $J_2=0.5$ &   $J_2=0.8$ &
\\ \hline\hline
& 111 & NC & -39.9516 &   -36.4735  &  -38.8786&\\
&&  NA & 0.0089  & 0.0104 & 0.0135&\\
\hline
& 222 & -39.3708 &   -39.9014  &  \textbf{-37.4680}  &  -40.4091  &\\
&&  0.0103 & 0.0113 & 0.0024 & 0.0090& \\ \hline
& 333 & \textbf{-43.8503}  & \textbf{ -39.9616}  &  -30.5563 &   -40.2787 &\\
&&  0.0078 & 0.0050 & 0.0177 & 0.0089 &\\ \hline
& 444 & -43.6545  &  -39.9399  &  -37.4598  &  \textbf{-41.5246} &\\
&& 0.0072 &  0.0088 & 0.0032 & 0.0104&\\ \hline
& 555 & -41.8887 &   -39.8155  &  -37.0180  &  -39.8954 &\\
&& 0.0152 & 0.0075  & 0.0024 &  0.0132 &\\ \hline
& DMRG &  -44.1277 & -40.7388 &   -37.5000 &   -42.0701 &
\\ \hline
\end{tabular}
\caption{The full results of the Heisenberg $J_1J_2$ VMC experiments with $J_1=1.0$ and different $J_2$ couplings, corresponding to different random seeds for  Euclidean GRU NQS. For each entry the energy is listed first in the first line, followed by the standard error in the second row. The best results are noted in bold. Non-converged results are marked as `NC' with the corresponding standard error marked as `NA'. }\label{full-j1j2-vmc-egru}
\end{table}
\newpage
\begin{table}[!h]
\centering
\begin{tabular}{c ccccc c}
\hline\hline
 &  Seed  & $J_2=0.0$ &   $J_2=0.2$  &   $J_2=0.5$ &   $J_2=0.8$ &
\\ \hline\hline
& 111 & \textbf{-42.5914} & \textbf{-39.6020}  & NC &  -33.6460 &\\
&& 0.0141  & 0.0100  & NA & 0.0344  &\\
\hline
& 222 &  -42.1332  & -39.5921  &  -35.8814  &  NC &\\
&& 0.0177  &  0.0113 &  0.0083 & NA & \\ \hline
& 333 &  -40.7290  &  -38.1450  & NC  & -36.7689&\\
&&  0.0239 & 0.0205  &NA &  0.0287 &\\ \hline
& 444 & -42.0836  &  -39.5835 &   NC & NC &\\
&&0.0181 &   0.0116  & NA & NA &\\ \hline
& 555 & -42.1118  &  -39.2110   & \textbf{-35.9253}   & \textbf{-38.4681} &\\
&&0.0185  &  0.0138 & 0.0078 &  0.0184 & \\ \hline
& DMRG &  -44.1277 & -40.7388 &   -37.5000 &   -42.0701 &
\\ \hline
\end{tabular}
\caption{The full results of the Heisenberg $J_1J_2$ VMC experiments with $J_1=1.0$ and different $J_2$ couplings, corresponding to different random seeds for Poincar\'e RNN NQS. For each entry the energy is listed first in the first line, followed by the standard error in the second row. The best results are noted in bold. Non-converged results are marked as `NC' with the corresponding standard error marked as `NA'. }\label{full-j1j2-vmc-prnn}
\end{table}
\begin{table}[!h]
\centering
\begin{tabular}{c ccccc c}
\hline\hline
 &  Seed  & $J_2=0.0$ &   $J_2=0.2$  &   $J_2=0.5$ &   $J_2=0.8$ &
\\ \hline\hline
& 111 & -43.0732   & \textbf{-40.4190}  &  \textbf{-37.4861}   & -41.4282 &\\
&&  0.0141 & 0.0053 & 0.0018 & 0.0070 &\\
\hline
& 222 & -42.8456 &   -40.2602  &  -37.4815 &   -40.7041  &\\
&& 0.0154 &  0.0052 & 0.0017 & 0.0137 & \\ \hline
& 333 & -43.0718  &  -40.1020  &  -37.4230  &  -39.2567 &\\
&&  0.0089 &  0.0045&  0.0039 & 0.0166 &\\ \hline
& 444 &  \textbf{-43.8982}  & -39.6266  &  -37.3639  &  \textbf{-41.6771}&\\
&&0.0079  &  0.0105 & 0.0046 & 0.0044 &\\ \hline
& 555 & -42.9658  &  -39.7689  &  -36.5931  &  -41.4085&\\
&& 0.0122 &  0.0066 & 0.0064&  0.0086& \\ \hline
& DMRG &  -44.1277 & -40.7388 &   -37.5000 &   -42.0701 &
\\ \hline
\end{tabular}
\caption{The full results of the Heisenberg $J_1J_2$ VMC experiments with $J_1=1.0$ and different $J_2$ couplings, corresponding to different random seeds for Poincar\'e GRU NQS. For each entry the energy is listed first in the first line, followed by the standard error in the second row. The best results are noted in bold.  }\label{full-j1j2-vmc-pgru}
\end{table}

\begin{table}[!h]
\centering
\begin{tabular}{c ccccc c}
\hline\hline
 &  Seed  & $J_2=0.0$ &   $J_2=0.2$  &   $J_2=0.5$ &   $J_2=0.8$ &
\\ \hline\hline
& 111 & \textbf{-44.0144}  &  -40.3537   & -36.8671   & \textbf{-40.1873} &\\
&& 0.0049 & 0.0051 &  0.0124 & 0.0154  &\\
\hline
& 222 &NC &  -40.1866   & -37.0814 &   -37.3330  &\\
&&NA &    0.0092 & 0.0134 & 0.0110  & \\ \hline
& 333 & NC &  \textbf{-40.5609}  & -35.6058  & NC &\\
&&  NA & 0.0047 & 0.0165 & NA &\\ \hline
& 444 & -42.2716 & -40.4646 & -36.8602 & -36.9563 &\\
&&0.0153 & 0.0058 &  0.0078 & 0.0244 &\\ \hline
& 555 & -43.2018  &  NC & \textbf{-37.1428}   & NC&\\
&& 0.0031 & NA & 0.0046 & NA & \\ \hline
& DMRG &  -44.1277 & -40.7388 &   -37.5000 &   -42.0701 &
\\ \hline
\end{tabular}
\caption{The full results of the Heisenberg $J_1J_2$ VMC experiments with $J_1=1.0$ and different $J_2$ couplings, corresponding to different random seeds for Lorentz RNN NQS. For each entry the energy is listed first in the first line, followed by the standard error in the second row. The best results are noted in bold. Non-converged results are marked as `NC' with the corresponding standard error marked as `NA'. }\label{full-j1j2-vmc-lrnn}
\end{table}
\begin{table}[!h]
\centering
\begin{tabular}{c ccccc c}
\hline\hline
 &  Seed  & $J_2=0.0$ &   $J_2=0.2$  &   $J_2=0.5$ &   $J_2=0.8$ &
\\ \hline\hline
& 111 &-43.8157 & -40.0406  &  \textbf{-37.4890} &   -40.2897 &\\
&&  0.0049 &  0.0055  & 0.0016  & 0.0129 &\\
\hline
& 222 & -43.2119 &  -39.6045   & -36.9744 &   -40.3649  &\\
&&  0.0067 &  0.0095 & 0.0026 & 0.0141& \\ \hline
& 333 & -42.7446  &  -39.3918  &  -36.8956  & NC  &\\
&& 0.0083 &  0.0047 & 0.0054 & NA  &\\ \hline
& 444 & -43.7173 &  -40.2761  &  NC & -39.8248 &\\
&&0.0047  & 0.0037 & NA & 0.0131 &\\ \hline
& 555 & \textbf{-43.9882} & \textbf{-40.3153}  & NC  & \textbf{-41.5325} &\\
&& 0.0038  & 0.0086 & NA &0.0028 & \\ \hline
& DMRG &  -44.1277 & -40.7388 &   -37.5000 &   -42.0701 &
\\ \hline
\end{tabular}
\caption{The full results of the Heisenberg $J_1J_2$ VMC experiments with $J_1=1.0$ and different $J_2$ couplings, corresponding to different random seeds for Lorentz GRU NQS. For each entry the energy is listed first in the first line, followed by the standard error in the second row. The best results are noted in bold. Non-converged results are marked as `NC' with the corresponding standard error marked as `NA'. }\label{full-j1j2-vmc-lgru}
\end{table}

\newpage
\subsubsection{$J_1J_2J_3$ experiments} \label{full-res-j1j2j3}
In this section, we list the full results of all individual Heisenberg $J_1J_2J_3$ VMC experiment run where one run corresponds to a single value of the $(J_2, J_3)$ coupling pair with a single NQS ansatz at a fixed random seed. In a manner identical to the $J_1J_2$ case above, for each NQS ansatz at a fixed $(J_2, J_3)$ pair, we performed at least 5 runs with 5 different random seeds (111, 222, 333, 444, 555) while keeping all other hyperparameters such as Euclidean and hyperbolic learning rates, as well as spatial constraint hyperparameters fixed. To select suitable hyperparameters for each ansatz, the 5-seed experiment runs have to be repeated with different ranges of the hyperparameters to tune. In what follows, we only include the chosen results obtained with the optimally selected hyperparameters.  In some of the runs, for all NQS ans\"atze including both Euclidean and hyperbolic variants, we obtained non-converged results where the mean energy produced by the  NQS ansatz drifted around without settling into a recognizable minimum or the convergence was at a false minimum that is much higher (e.g. positive energy) than the actual true ground state energy. These non-converged results were not taken into account when calculating the averaged values reported in Table \ref{j1j2j3-RNN-GRU-ave}.
\begin{table}[!h]
\centering
\begin{tabular}{c ccccc c}
\hline\hline
 &  Seed  & (0.0,0.5) &   (0.2,0.2) &   (0.2,0.5) &   (0.5,0.2) &
\\ \hline\hline               
& 111 & -37.5211   & -25.3339 &   -32.5270  &  -17.5030 &\\
&&  0.0011 & 0.0024 & 0.0006 &  0.0009 &\\
\hline
& 222 & \textbf{-37.6130}  &  \textbf{-25.3654}  &  -32.6383 &    -17.5000&\\
&& 0.0022  & 0.0009  & 0.0009  & 0.0010 & \\ \hline
& 333 & -37.4966   & -25.1499   &  -32.5273  &  -17.4917 &\\
&&  0.0027 &  0.0033 &  0.0006 & 0.0058 &\\ \hline
& 444 & -35.5873   & -25.2236   & -31.0007   & -17.8513 &\\
&& 0.0018  & 0.0044 &  0.0004 &  0.0032 &\\ \hline
& 555 &-37.5199  &  -25.1532 &   \textbf{-32.6716}   & \textbf{-18.0043} &\\
&& 0.0011 & 0.0113 & 0.0052 &  0.0013 & \\ \hline
& DMRG &    -53.9914   &-43.5860 &   -49.6287   &  -38.5473 &
\\ \hline
\end{tabular}
\caption{The full results of the Heisenberg $J_1J_2J_3$ VMC experiments with $J_1=1.0$ and different $(J_2, J_3)$ couplings, corresponding to different random seeds for Euclidean RNN NQS. For each entry the energy is listed first in the first line, followed by the standard error in the second row. The best results are noted in bold. }\label{full-j1j2j3-vmc-ernn}
\end{table}
\begin{table}[!h]
\centering
\begin{tabular}{c ccccc c}
\hline\hline
 &  Seed  & (0.0,0.5) &   (0.2,0.2) &   (0.2,0.5) &   (0.5,0.2) &
\\ \hline\hline               
& 111 &-38.2468 & -40.9102  &  -48.7788  &  \textbf{-37.7999} &\\
&& 0.0445  & 0.0078 & 0.0159 & 0.0278  &\\
\hline
& 222 & -37.5188 &  -33.2060 &   \textbf{-48.9555}  &  -37.2210  &\\
&& 0.0069 &   0.0116 & 0.0140&  0.0070 & \\ \hline
& 333 & NC &  \textbf{-42.8726} & -48.4612 &  -37.5307 &\\
&&  NA &  0.0059 &  0.0193 & 0.0111 &\\ \hline
& 444 & \textbf{-53.2965}  &  -41.3130 &   -48.8926  &  NC &\\
&& 0.0161 &  0.0095 &  0.0158  & NA &\\ \hline
& 555 &-37.6151 &-42.0390  &  -48.3670  &  -37.2893 &\\
&& 0.0076 & 0.0155 & 0.0099 & 0.0077& \\ \hline
& DMRG &    -53.9914   &-43.5860 &   -49.6287   &  -38.5473 &
\\ \hline
\end{tabular}
\caption{The full results of the Heisenberg $J_1J_2J_3$ VMC experiments with $J_1=1.0$ and different $(J_2, J_3)$ couplings, corresponding to different random seeds for  Euclidean GRU NQS. For each entry the energy is listed first in the first line, followed by the standard error in the second row. The best results are noted in bold. Non-converged results are marked as `NC' with the corresponding standard error marked as `NA'. }\label{full-j1j2j3-vmc-egru}
\end{table}

\begin{table}[!h]
\centering
\begin{tabular}{c ccccc c}
\hline\hline
 &  Seed  & (0.0,0.5) &   (0.2,0.2) &   (0.2,0.5) &   (0.5,0.2) &
\\ \hline\hline               
& 111 & -50.6413  &  -41.5643   & -46.5458 &   -36.9627&\\
&&  0.0292 & 0.0178 &  0.0254 & 0.0152 &\\
\hline
& 222 &  -50.7754 &  -41.6428 &  \textbf{-46.8415} &    NC &\\
&&0.0276  &  0.0182 &  0.0250 &  NA& \\ \hline
& 333 & \textbf{-50.9784} &   \textbf{-41.7109}  &   -46.2106 &   \textbf{-37.2335} &\\
&&  0.0270 & 0.0177&  0.0270&  0.0090 &\\ \hline
& 444 & -50.8889  &  -41.6916 &   -45.3887  &  -36.9736 &\\
&& 0.0263  & 0.0177&  0.0293 & 0.0148 &\\ \hline
& 555 & -50.5952 &   -41.6616   & -46.7605  &  -36.6943 &\\
&& 0.0273  & 0.0190 &  0.0255  & 0.0185& \\ \hline
& DMRG &    -53.9914   &-43.5860 &   -49.6287   &  -38.5473 &
\\ \hline
\end{tabular}
\caption{The full results of the Heisenberg $J_1J_2J_3$ VMC experiments with $J_1=1.0$ and different $(J_2, J_3)$ couplings, corresponding to different random seeds for Poincar\'e RNN NQS. For each entry the energy is listed first in the first line, followed by the standard error in the second row. The best results are noted in bold. Non-converged results are marked as `NC' with the corresponding standard error marked as `NA'. }\label{full-j1j2j3-vmc-prnn}
\end{table}

\begin{table}[!h]
\centering
\begin{tabular}{c ccccc c}
\hline\hline
 &  Seed  & (0.0,0.5) &   (0.2,0.2) &   (0.2,0.5) &   (0.5,0.2) &
\\ \hline\hline               
& 111 &-52.8287 & \textbf{-42.8895}  &  \textbf{-48.9895}   & \textbf{-38.3142}&\\
&& 0.0383 &  0.0083 & 0.0145 &  0.0052  &\\
\hline
& 222 & -52.6683  &  -42.3495  &  -47.8148  &  NC  &\\
&&0.0159  &  0.0079 &  0.0183  & NA & \\ \hline
& 333 & \textbf{-53.5432} & -42.4183 & -48.3456  & -36.7277 &\\
&& 0.0201 &  0.0120 &  0.0171 &  0.0193  &\\ \hline
& 444 & -53.4281  & -42.7937  & -48.9116 & -36.5693 &\\
&& 0.0119 &  0.0113 & 0.0089  & 0.0171&\\ \hline
& 555 &-53.4456 & -42.2968  & -47.8189  & -37.3046 &\\
&&0.0134 & 0.0093 & 0.0097 & 0.0163 & \\ \hline
& DMRG &    -53.9914   &-43.5860 &   -49.6287   &  -38.5473 &
\\ \hline
\end{tabular}
\caption{The full results of the Heisenberg $J_1J_2J_3$ VMC experiments with $J_1=1.0$ and different $(J_2, J_3)$ couplings, corresponding to different random seeds for Poincar\'e GRU NQS. For each entry the energy is listed first in the first line, followed by the standard error in the second row. The best results are noted in bold. Non-converged results are marked as `NC' with the corresponding standard error marked as `NA'. }\label{full-j1j2j3-vmc-pgru}
\end{table}
\begin{table}[!h]
\centering
\begin{tabular}{c ccccc c}
\hline\hline
 &  Seed  & (0.0,0.5) &   (0.2,0.2) &   (0.2,0.5) &   (0.5,0.2) &
\\ \hline\hline               
& 111 & -53.7343  &  -43.4213  &  -49.1737  &  -38.2994 &\\
&&0.0112 &   0.0045 & 0.0114 & 0.0074&\\
\hline
& 222 &  -53.6295  & NC &  -49.5429   & NC &\\
&& 0.0173 & NA &  0.0126 & NA & \\ \hline
& 333 & NC  & NC &  \textbf{-49.5567}  &  NC &\\
&&  NA&  NA&  0.0083  & NA &\\ \hline
& 444 &  NC &  \textbf{-43.4314}  & NC  & NC &\\
&& NA &   0.0058 & NA &   NA & \\ \hline
& 555 & \textbf{-53.9317} & NC &  -49.4750   &  \textbf{-38.3244} &\\
&&0.0306   & NA & 0.0084 &  0.0042 & \\ \hline
& DMRG &    -53.9914   &-43.5860 &   -49.6287   &  -38.5473 &
\\ \hline
\end{tabular}
\caption{The full results of the Heisenberg $J_1J_2J_3$ VMC experiments with $J_1=1.0$ and different $(J_2, J_3)$ couplings, corresponding to different random seeds for Lorentz RNN NQS. For each entry the energy is listed first in the first line, followed by the standard error in the second row. The best results are noted in bold. Non-converged results are marked as `NC' with the corresponding standard error marked as `NA'. }\label{full-j1j2j3-vmc-lrnn}
\end{table}
\begin{table}[!h]
\centering
\begin{tabular}{c ccccc c}
\hline\hline
 &  Seed  & (0.0,0.5) &   (0.2,0.2) &   (0.2,0.5) &   (0.5,0.2) &
\\ \hline\hline               
& 111 & \textbf{-53.8242} & -42.7053 &    \textbf{-49.1531} &   -37.7468 &\\
&& 0.0094 &  0.0067 & 0.0073  & 0.0062  &\\
\hline
& 222 &  NC &  NC &  -48.7920 &  NC &\\
&& NA &    NA &  0.0118 &  NA & \\ \hline
& 333 & NC  & -41.9637   & NC  & -37.4020  &\\
&& NA &   0.0065 & NA & 0.0025  &\\ \hline
& 444 & -50.0773 &  -42.9040 &   -46.4709   &  \textbf{-38.0665} &\\
&& 0.0253 &  0.0062 & 0.0220 & 0.0044&\\ \hline
& 555 & -53.7332  &   \textbf{-43.1327}  &  -48.8566   & -37.8918&\\
&&0.0106 &   0.0084  &0.0095 & 0.0054 & \\ \hline
& DMRG &    -53.9914   &-43.5860 &   -49.6287   &  -38.5473 &
\\ \hline
\end{tabular}
\caption{The full results of the Heisenberg $J_1J_2J_3$ VMC experiments with $J_1=1.0$ and different $(J_2, J_3)$ couplings, corresponding to different random seeds for  Lorentz GRU NQS. For each entry the energy is listed first in the first line, followed by the standard error in the second row. The best results are noted in bold. Non-converged results are marked as `NC' with the corresponding standard error marked as `NA'. }\label{full-j1j2j3-vmc-lgru}
\end{table}


\end{document}